\documentclass[letterpaper,12pt]{article}   
\usepackage{osajnl2} 
\usepackage[draft]{hyperref} 
\usepackage{subfigure}
\usepackage{amsmath}
\usepackage{booktabs}

\newcommand{\C}{\mathcal C}
\newcommand{\eye}{\mathcal I}
\newcommand{\e}[1]{\times 10^{#1}} 
\newcommand{\sinc}{\, \textrm{sinc}} 
 
\newcommand{\fig}[1]{Fig.~\ref{#1}} 
\newcommand{\eq}[1]{Eq.~\ref{#1}}

\begin{document}

\title{Kalman filtering techniques for focal plane electric field estimation}


\author{Tyler D. Groff,$^{1,*}$ N. Jeremy Kasdin,$^1$}
\address{$^1$Depatment of MAE, EQuad, Olden St \\ Princeton, NJ 08544, USA}
\address{$^*$Corresponding author: tgroff@princeton.edu}

\begin{abstract}For a coronagraph to detect faint exoplanets, it will require focal plane wavefront control techniques to continue reaching smaller angular separations and higher contrast levels. These correction algorithms are iterative and the control methods need an estimate of the electric field at the science camera, which requires nearly all of the images taken for the correction. The best way to make such algorithms the least disruptive to science exposures is to reduce the number required to estimate the field. We demonstrate a Kalman filter estimator that uses prior knowledge to create the estimate of the electric field, dramatically reducing the number of exposures required to estimate the image plane electric field while stabilizing the suppression against poor signal-to-noise (SNR). In addition to a significant reduction in exposures, we discuss the relative merit of this algorithm to estimation schemes that do not incorporate prior state estimate history, particularly in regard to estimate error and covariance. Ultimately the filter will lead to an adaptive algorithm which can estimate physical parameters in the laboratory for robustness to variance in the optical train.
\end{abstract}


\maketitle 

\section{Introduction}\label{sec:intro}
The desire to directly image extrasolar terrestrial planets has motivated much research into space-based missions. One approach proposed for direct imaging in visible to near-infrared light is a coronagraph, which use internal masks and stops to change the point spread function of the telescope, creating regions in the image of high contrast where a dim planet can be seen.  However, coronagraphs possess an extreme sensitivity  to wavefront aberrations generated by the errors in the system optics. This necessitates wavefront control algorithms to correct for the aberrations and relax manufacturing tolerances and stability requirements within the observatory. Of the two, stability is of particular importance as noted by Shaklan \cite{shaklan2006reflectivity,shaklan2011stability}. In this paper we discuss the challenges associated with wavefront estimation and control using deformable mirrors (DMs) in a coronagraphic imager, and how we can improve the robustness of such a system to observatory instability. Advances in these correction algorithms have primarily focused on development of the controller, by choosing some criterion that decides how best to suppress aberrations given an estimate of the electric field at that point in time. This estimate is found by modulating the system in a predicable fashion so that there is adequate phase modulation to solve for the expected value of the electric field at the image plane. Whether this modulation is from observations at multiple planes, or by modulating the the surface of a DM, the observations are made in the presence of the all prior control signals sent to the DMs. Since existing estimation schemes do not predict the effect of a perturbation induced by the DM they must abandon all prior knowledge of the electric field estimate, and as such require a large number of images to reconstruct the estimate. By utilizing both prior estimates and the control history we develop a method that requires fewer images to update the estimate, simultaneously improving the efficiency of the correction algorithm and its robustness to observatory instability.

\section{Experimental Setup: Princeton High Contrast Imaging Laboratory}\label{HCIL}

\begin{figure}[ht!]
\centering
\includegraphics[width = .55\textwidth]{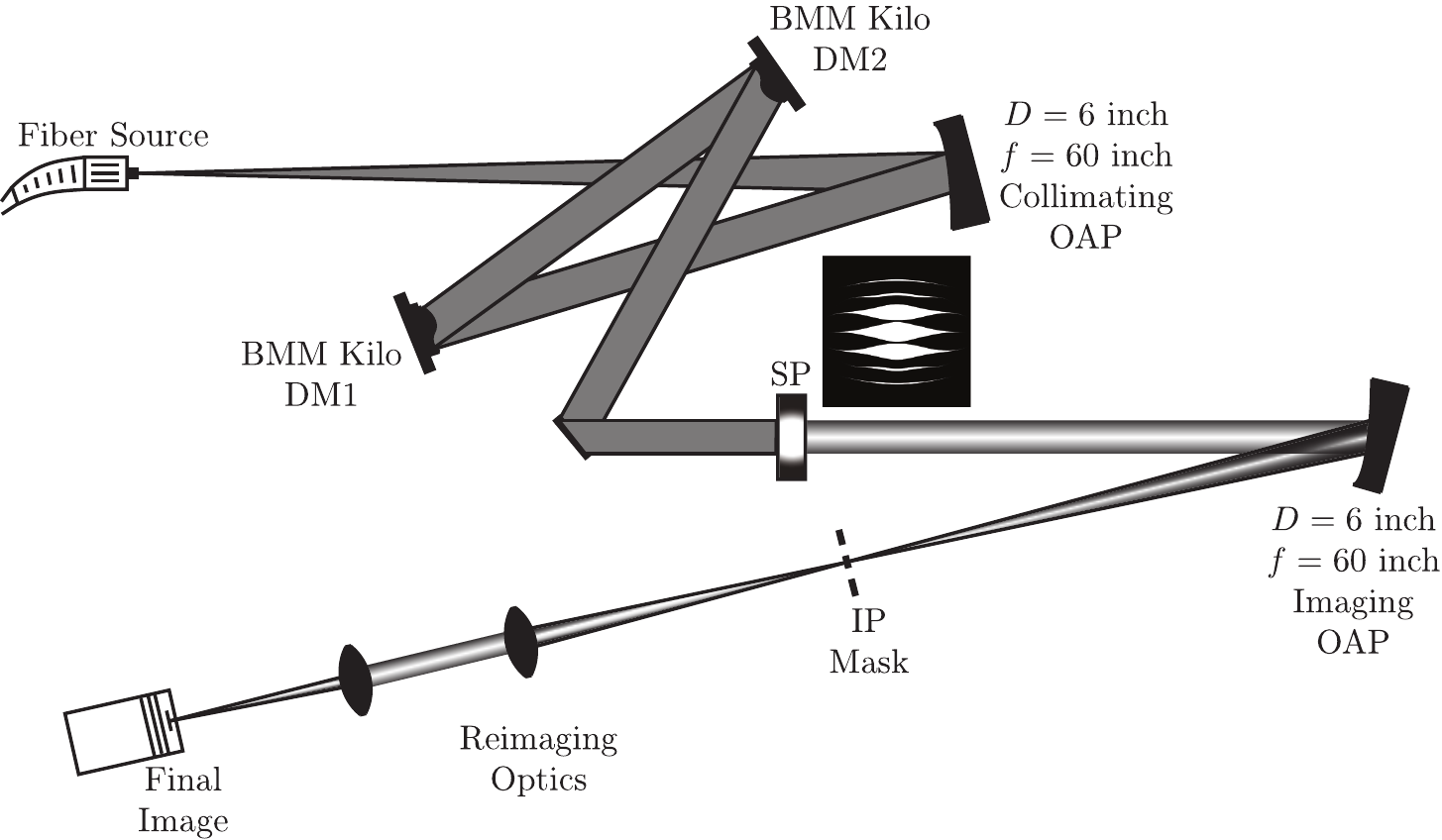}
\caption{Optical Layout of the Princeton HCIL.  Collimated light is incident on two DMs in series, which propagates through a Shaped Pupil, the core of the PSF is removed with an image plane mask, and the $90^\circ$ search areas are reimaged on the final camera.}
\label{optical}
\end{figure}

\begin{figure}[ht!]
\centering
\subfigure[Shaped Pupil] {
    \label{SP}
    \includegraphics[width = .205\columnwidth,clip=true,trim=.05in 0in .35in 0in]{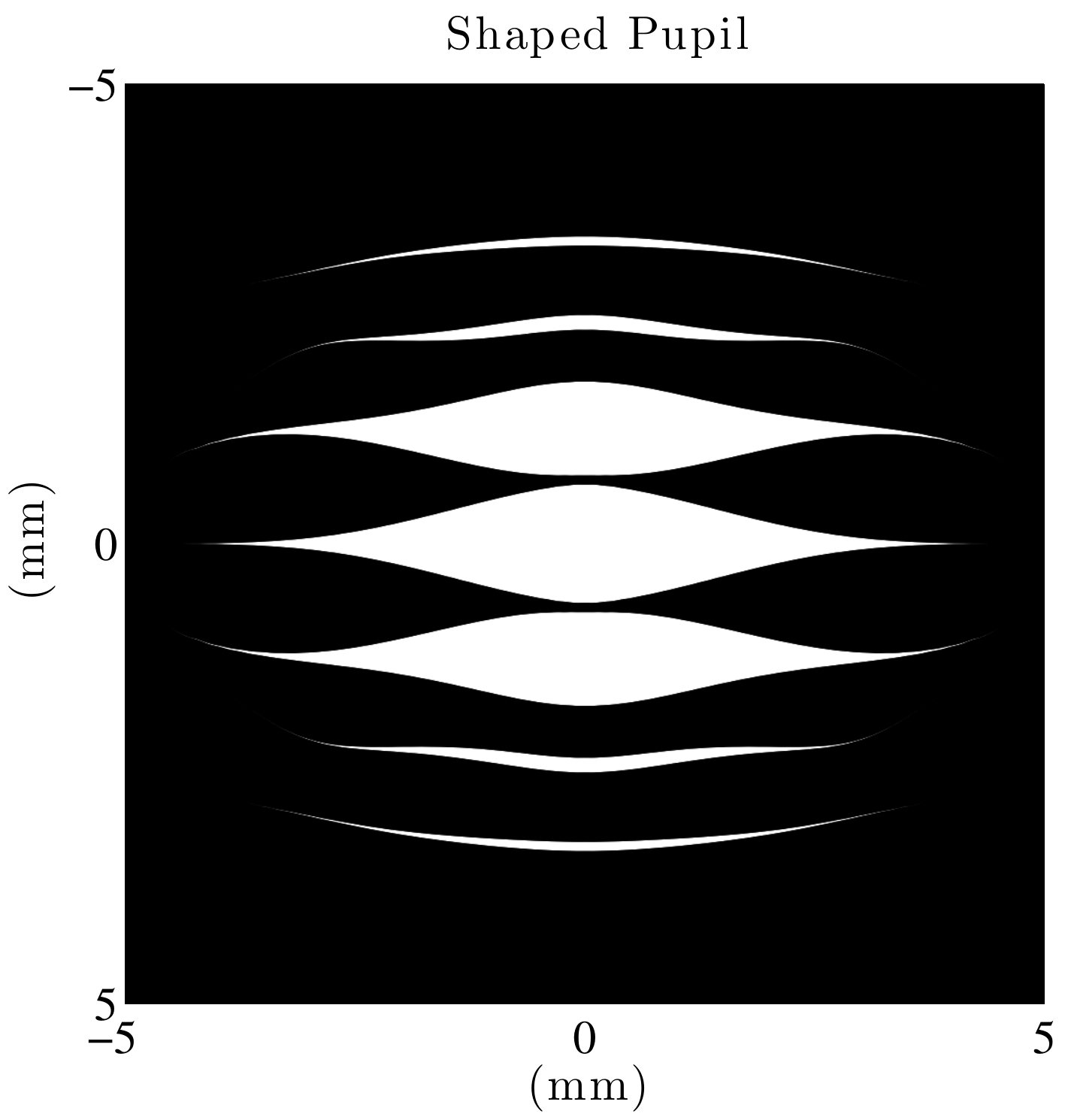}}
\subfigure[Ideal PSF] {
    \label{Ideal_PSF}
    \includegraphics[width = .24\columnwidth,clip=true,trim=.05in 0in .35in 0in]{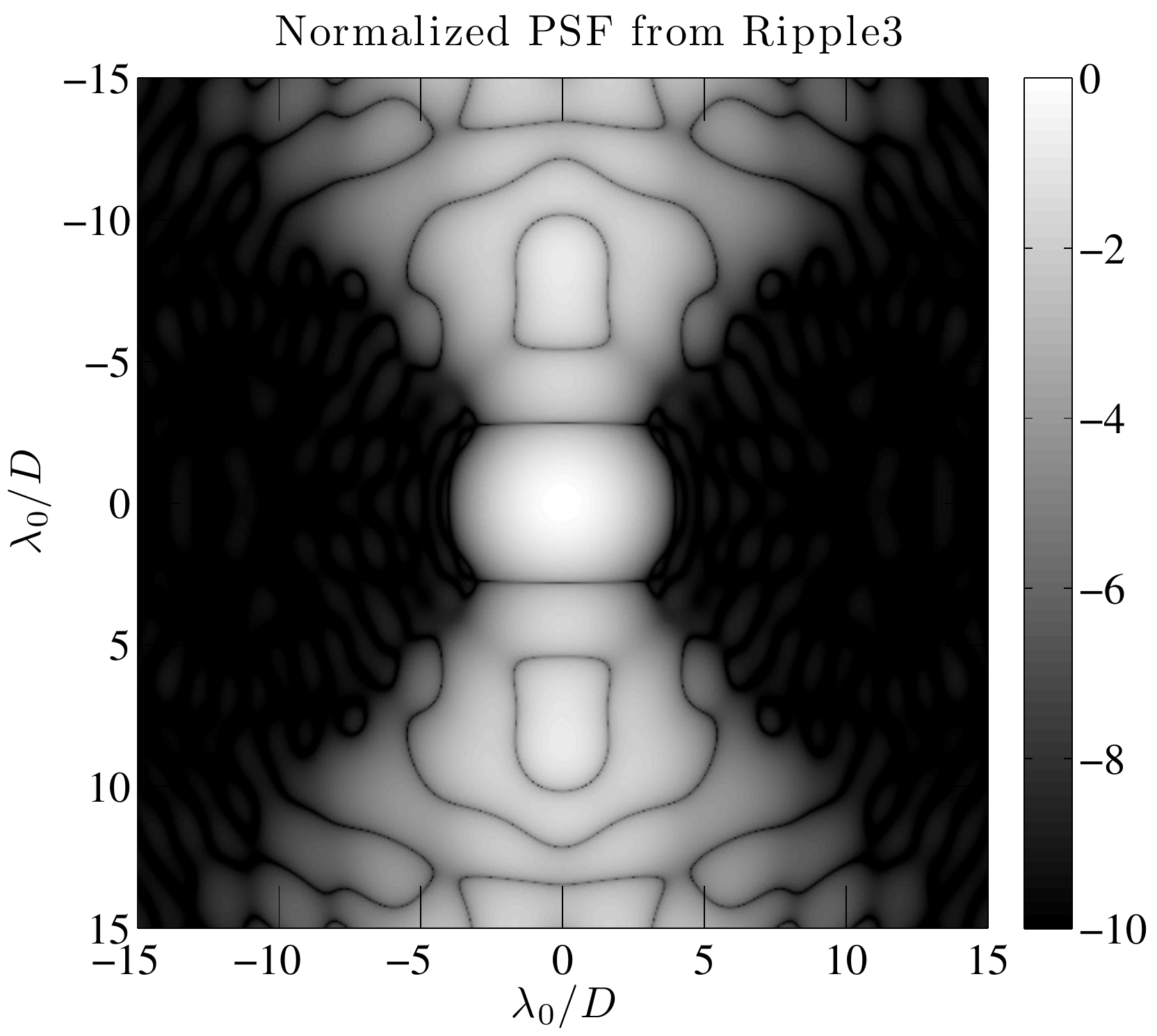}}\\
   \subfigure[Aberrated Pupil]{
    \label{ab_SP}
    \includegraphics[width = .205\columnwidth,clip=true,trim=.05in 0in .35in 0in]{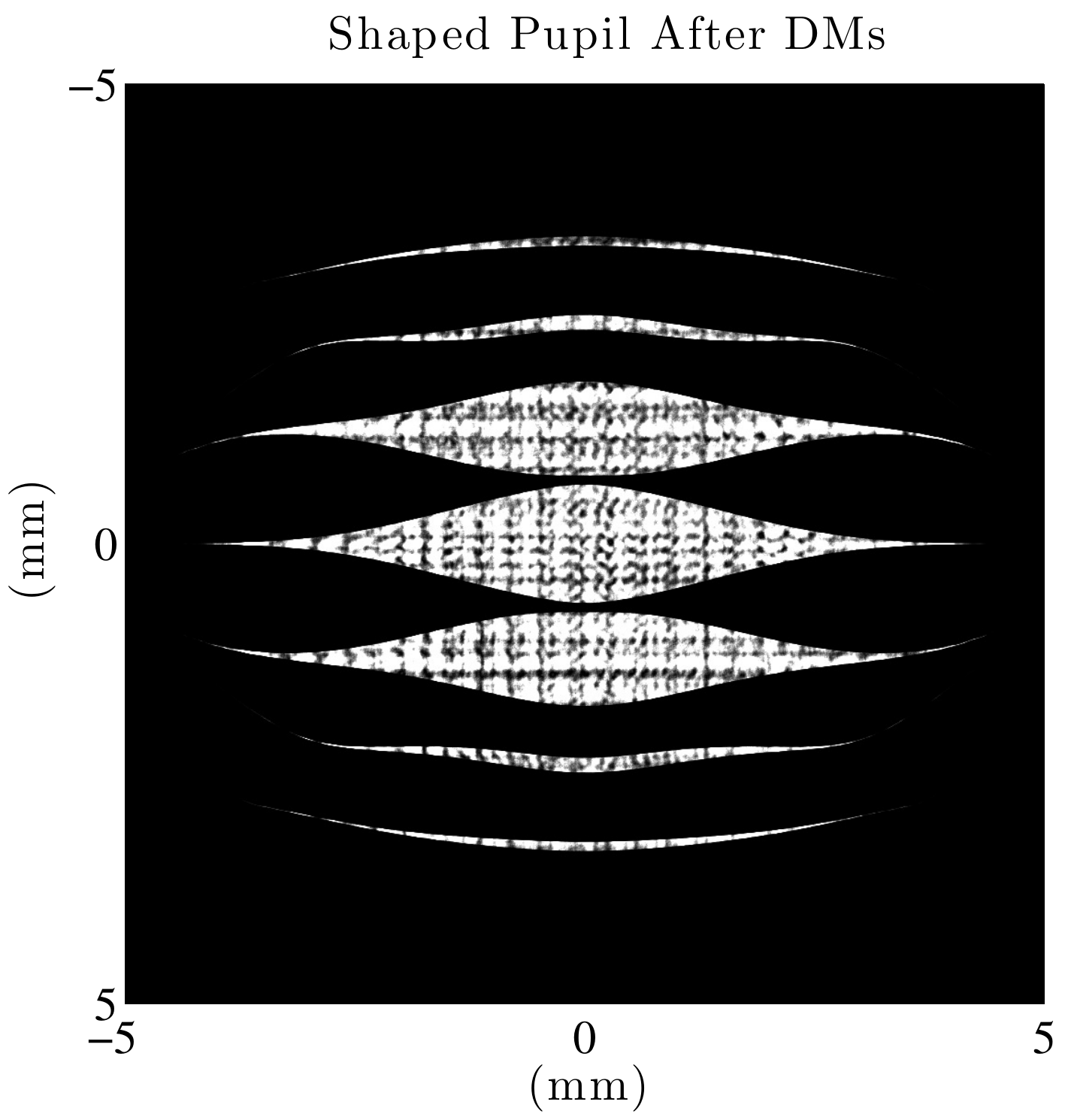}\label{SPab}}
    \subfigure[Aberrated PSF]{
    \label{ab_PSF}
    \includegraphics[width = .24\columnwidth,clip=true,trim=.05in 0in .35in 0in]{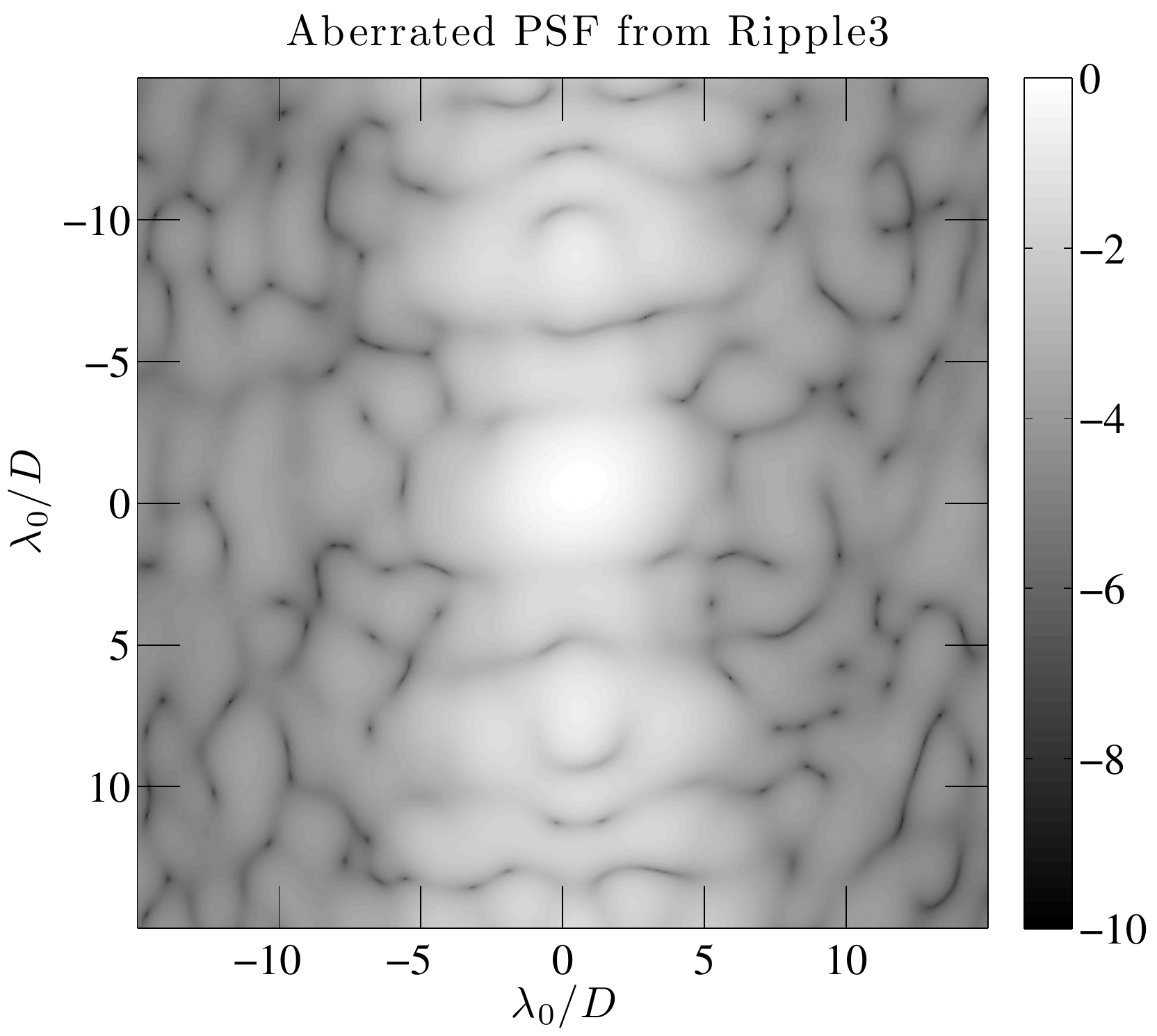}}
    \caption{(a) A shaped pupil.  (b)  The ideal PSF from a system using a shaped-pupil coronagraph.  (c)  Shaped Pupil with aberrations generated by Fresnel propagating the measured nominal shapes of the DMs to the pupil plane. Other sources of aberrations are not included because they have not been measured. (d) The PSF of the shaped pupil with the simulated aberrations.   The figures are in a log scale, and the log of contrast is shown in the colorbars.}
    \label{image}
\end{figure}

The High Contrast Imaging Laboratory (HCIL) at Princeton tests coronagraphs and wavefront control algorithms for quasi-static speckle suppression. The collimating optic is a six inch off-axis parabola (OAP) followed by two DMs in series and a coronagraph, which is imaged with a second six inch OAP (\fig{optical}).  We use a shaped pupil coronagraph, shown in \fig{SP}, and described in detail in Belikov et al.\cite{belikov2007broadband}.  This coronagraph produces a discovery space with a theoretical contrast of $3.3\times10^{-10}$ in two $90^\circ$ regions as shown in \fig{Ideal_PSF}. At the Princeton HCIL, the  aberrations in the system result in an uncorrected average contrast of $\sim1\times10^{-4}$ in the area immediately surrounding the core of the point spread function (PSF), which agrees with the simulations shown in \fig{ab_PSF}. Since the coronagraph is a binary mask, its contrast performance is fundamentally achromatic, subject only to the physical scaling of the PSF with wavelength.

\section{Modeling the HCIL}\label{HCILmodel}
The estimation and control algorithms in this paper are dependent on a model of the optical system, specifically the transfer function between the DM and image plane. We detail a common approach to this mapping, where the field is linearized to describe the problem as discrete perturbations beginning at the pupil plane \cite{malbet1995high,trauger2004coronagraph,giveon2007broadband,GBSK07,pueyo2009optimal,guyon2009labvalid}. Most commonly, the field is linearized about the current DM surface shape. In the Princeton HCIL neither DM is conjugate to the pupil plane, requiring that we account for this propagation in our  model. Beginning with arbitrary complex aberrated field, $g(u,v)$, incident on arbitrary aperture, $A(u,v)$, and a phase perturbation induced by the DM surface, $\tilde \phi(u,v)$, the field at the DM plane is given by
\begin{equation}
E_{0}(u,v) = A(u,v) (1 + g(u,v)) e^{i \tilde \phi(u,v)}.
\end{equation}
The DM perturbation commanded by the controller, $\phi(u,v)$, will be added to the prior (also referred to as ``nominal") DM shape, $\phi_0(u,v)$, so that $\tilde \phi = \phi + \phi_0$. Linearizing about the nominal DM shape, the field at the pupil plane is given by
\begin{equation}
E_{0}(u,v) \cong A(u,v) e^{i \phi_0}(1 + g(u,v) + i \phi(u,v)). \label{eq:linpupnonzero}
\end{equation}

The resultant image plane electric field, $E_{im}(x,y)$, is described by an arbitrary linear operator, $\C\{\cdot\}$, as
\begin{align}
E_{im}(x,y) &= \C\{E_{0}(u,v)\}\\
&= \C\{A(u,v) e^{i \phi_0}\} + \C\{A(u,v) e^{i \phi_0}g(u,v)\} + i \C\{ A(u,v) e^{i \phi_0} \phi(u,v)\}.
\end{align}
The intensity distribution of this field is given by 
\begin{equation}
I_{im}(x,y) = \biggl|  \C\{A(u,v) e^{i \phi_0}\} + \C\{A(u,v) e^{i \phi_0}g(u,v)\} + i \C\{ A(u,v) e^{i \phi_0} \phi(u,v)\}. \biggr|^2.\label{eq:intensitydef}
\end{equation}
Rewriting \eq{eq:intensitydef} as an inner product and absorbing $e^{i \phi_0}$ into $\C\{\cdot\}$, the intensity distribution in the image plane is represented as
\begin{equation}
\begin{split}
I_{im} = <\C\{A\phi\},\C\{A\phi\}> +& 2\Re\{<\C\{A(1+g)\},i \C\{A\phi\}>\} \\+& < \C\{A(1+g)\}, \C\{A(1+g)\}>.
\end{split} \label{eq:vectoridh}
\end{equation}
It is important to note that to re-linearize about the current DM shape at any control step, $k$, $\C\{\cdot\}$ must be re-evaluated. Fortunately, this is as simple as convolving $\C\{e^{i \phi_{k-1}}\}$ with each component of \eq{eq:vectoridh}. 

For the estimator we impose a scalar inner product to each element in the two-dimensional plane, $(x,y)$, to compute the intensity distribution across the image plane. For the control algorithm we seek an average contrast value inside the dark hole as a metric for the control law. Thus, we impose a matrix inner product to describe the image plane electric field, $\C\{Ag\}$, and the DM control effect, $\C\{A\phi\}$, our state and control variables respectively, as column matrices. These are formed by stacking the columns of the two-dimensional fields to create a single column matrix. However, we have yet to parameterize the DM commands into a control matrix, $u$. At the moment we could solve for $\C\{A\phi\}$ but what we seek are actuation commands for the DM, not its field at the image plane. To solve for the actuator commands, we introduce a physical model of the DM to directly optimize the amplitude of each DM actuator. Letting $H(x,y)$ be the height of the DM surface, the resulting phase perturbation induced by the DM, $\phi(u,v)$, is 
\begin{equation}
\phi(u,v) = \frac{2\pi}{\lambda_0} H(x,y).
\end{equation}
For control, we wish to describe $H(x,y)$ as a combination of the two-dimensional height maps imposed by each actuator. Since we are using a DM with a continuous face sheet, the contribution of any actuator will be highly localized but still deforms the entire DM surface. As a result, we must describe the contribution of the $q^{th}$ actuator as a two-dimensional phase map over the entire plane of the DM surface, $h_q(x,y)$. The combination of all actuators is nonlinear, and very complicated to compute \cite{blain2010dm}, but in the presence of small aberrations our deformations will be small. Thus, we assume that superposition of actuators is valid and describe $H(x,y)$ as a sum of $h_q(x,y)$ over all actuators, $N_{act}$. The phase contribution of the DM is then given by
\begin{equation}
\phi(u,v)= \frac{2\pi}{\lambda_0} \sum_{q=1}^{N_{act}} h_q(u,v).
\end{equation}
Finally, we wish to make our control matrix, $u$, a column matrix made up of the control signal from each actuator, $u_q$. To do so we describe $h_q(u,v)$ as a characteristic shape with unitary amplitude, commonly referred to as an influence function, $f_q(u,v)$. To find $h_q(u,v)$ we simply multiply $f_q(u,v)$ by the control amplitude, $a_q$. Describing $h_q(u,v)$ with influence functions, the phase perturbation induced by the DM is
\begin{equation}
\phi(u,v) = \frac{2\pi}{\lambda_0} \sum_{q=1}^{N_{act}} a_q f_q(u,v), \label{eq:infdx}
\end{equation}
which sums the $q^{th}$ 2-D phase map, or influence function $f_q(u,v)$, for all $N_{act}$ actuators to reconstruct $\phi(u,v)$. The strength of each influence function is determined by $a_q$. 

Substituting Eq.~\ref{eq:infdx} into Eq.~\ref{eq:vectoridh}, we write $\C\{A\phi\} = \C\{Af\} u$, where $u$ is a column matrix of actuator strengths, $u = [a_1 \dots a_k]^T$, and 
$\C\{Af\}$ can be written as a matrix of dimension $N_{pix} \times N_{act}$. To simplify the notation we define this matrix as the control effect matrix, 
\begin{equation}
G = \C\{Af\} \label{eq:g} \in [N_{pix} \times N_{act}].
\end{equation}
This allows us to write
\begin{equation}
<\C\{A\phi\},\C\{A\phi\}> = u^TG^*Gu,
\end{equation}
where ${}^*$ denotes the conjugate transpose. Applying the matrix form of the control amplitudes to Eq.~\ref{eq:vectoridh}, the scalar value for the average intensity in the dark hole, $I_{DH}$, is given by
\begin{equation}
I_{DH} = \frac{4 \pi^2}{\lambda^2} u^T M u + \frac{4 \pi}{\lambda} u^T \Im\{b\} + d. \label{eq:quadidh}
\end{equation}
Where 
\begin{align}
M &= <\C\{Af\},\C\{Af\}> = G^*G \label{eq:monom}\\
b &= <\C\{A(1+g)\},\C\{Af\}> =  G^*  \C\{A(1+g)\}\label{eq:monob}\\
d &= <\C\{A(1+g)\},\C\{A(1+g)\}> =  \C\{A(1+g)\}^* \C\{A(1+g)\}.
\end{align}
Conceptually, $d$ is the column matrix of the intensity contribution from the aberrated field, $b$ is a matrix representing the interaction of the DM electric field with the aberrated field, and $M$ describes the additive contribution of the DM to the image plane intensity. Having represented $I_{DH}$ in a quadratic form we can use Eq.~\ref{eq:quadidh} to produce an optimal control strategy, but we must first estimate $\C\{A(1+g)\}$ to compute $b$ at each control step. Doing so optimally is the main topic of this paper. 
\section{Stroke Minimization in Monochromatic Light}\label{stroke}
In general, focal plane wavefront correction methods are broken down into a wavefront estimation step followed by a control step where the DM surface height is determined. In this paper we use the stroke minimizing algorithm described in Pueyo et al. \cite{pueyo2009optimal} as the controller to test the estimation schemes. Stroke minimization corrects the wavefront by minimizing the actuator stroke on the DMs subject to a target contrast value\cite{pueyo2009optimal}. Using a single deformable mirror to control the field, these algorithms are capable of correcting both amplitude and phase aberrations on a single side of the image plane. Pueyo et al. \cite{pueyo2009optimal} showed that, to first order, two DMs in series are capable of correcting both amplitude and phase aberrations symmetrically about the optical axis, doubling the discovery space in the image plane. Physically, this is due to the phase-to-amplitude mixing from propagating the field between non-conjugate planes (the first DM to the second). Expressing the actuator amplitudes of both DMs as a vector, $u$, the optimization problem is written as
 \begin{equation}
\begin{array}{ccc}
\mbox{minimize} & & \displaystyle \sum_{k=1}^N a_k^2 = u^Tu \medskip\\
\mbox{subject to} & & I_{DH}  \le 10^{-C}.
\end{array}
\end{equation}
Using the quadratic form for $I_{DH}$ in \eq{eq:quadidh}, and applying the constraint to the minimization via the Lagrange multiplier, $\mu$, the quadratic cost function is given by
\begin{equation}
J = u^T\left( \eye + \mu \frac{4 \pi^2}{\lambda^2} M\right)u + \mu \frac{4 \pi}{\lambda} u^T \Im\{b\} + \mu \left(d- 10^{-C}\right). \label{eq:monocost}\\
\end{equation}
The corresponding optimal command that minimizes \eq{eq:monocost} is given by
\begin{equation}
u_{opt} = - \mu \left(\frac{\lambda}{2\pi} \eye + \mu \frac{2\pi}{\lambda} M \right)^{-1} \Im\{b\}.\label{eq:monocontrol}
\end{equation}

We find the optimal actuator commands via a line search on $\mu$ to minimize the augmented cost function, \eq{eq:monocost}. Since this is a quadratic subprogram of the full nonlinear problem we can iterate to reach any target contrast \cite{pueyo2009optimal}. In addition to regularizing the problem of minimizing the contrast in the search area, minimizing the stroke has the added advantage of keeping the actuation small and thus within the linear approximation. If the DM model and its transformation to the electric field (embedded in the $M$ matrix) were perfectly known, the achievable monochromatic contrast  would be limited only by estimation error.  
Note that using the derivation shown in \S\ref{HCILmodel} we have only explicitly derived the control law for a single DM. However, Pueyo et al. \cite{pueyo2009optimal} have shown that $M$ and $b$ only need to be augmented to account for additional DMs at non-conjugate planes. Thus the form of the control law remains the same, and only the dimension of the matrices change.
\section{Least-Squares Estimation Using Pairwise Measurements}\label{batch}
To date, almost all high-contrast wavefront control approaches use the DM diversity algorithm developed by Give'on et al. \cite{giveon2007broadband} to estimate the wavefront. Up to this point DM diversity had given the best results at the Princeton HCIL in both monochromatic and broadband light \cite{pueyo2009optimal,groff2012aeroconf}, making it the baseline of comparison for the Kalman filter estimation scheme. The DM diversity algorithm solves for the electric field by modulating a single DM while measuring the variation in intensity at the image plane. We then solve for the field by constructing an observation matrix based on a model of how that DM perturbs the electric field at the image plane. 

The linearized interaction of the DM actuation with the aberrated electric field can be written in matrix form by taking difference images using $j$ pre-determined shapes with amplitudes prescribed by the normalized intensity of the aberrated field \cite{borde2006speckle,giveon2011pair}. The image $I_j^+$ is taken with one deformable mirror shape, $\phi_j$, while $I_j^-$ is the image taken with the negative of that shape, $-\phi_j$, applied to the deformable mirror. Applying \eq{eq:vectoridh}, the resulting measurement is given by
\begin{equation}
I^+ -I^- =  4 \Re\{<\C\{A(1+g)\},i \C\{A\phi\}>\}. \label{eq:diffim}
\end{equation}
Taking $j$ measurements, each with a different conjugate pair of DM shapes,  a vector of noisy measurements at the $k^{th}$ time step is constructed as 
\begin{equation}
z_k = \left[\begin{array}{c}I_{1,k}^+ - I_{1,k}^-\\ \dots \\ I_{j,k}^+ - I_{j,k}^-\end{array}\right]. \label{eq:z}
\end{equation}
We construct the observation matrix, $H_k$, so that it contains the real and imaginary parts of the $j^{th}$ DM perturbation, $C\{A\phi_j\}$, in each row. With $j$ DM probe shapes, the observation matrix at the $k^{th}$ time step is given by
\begin{equation}
H_k = 4 \left[\begin{array}{cc}\Re\{C\{A \phi_{1,k} \}\} & \Im\{C\{A \phi_{1,k} \}\}\\ \vdots&\vdots \\ \Re\{C\{A \phi_{j,k}\} \} & \Im\{C\{A \phi_{j,k}\} \} \end{array}\right].
\end{equation}
Having defined the form of $H_k$, it follows from \eq{eq:diffim} that the image plane electric field is separated into real and imaginary parts. The state vector describing the image plane electric field of a specific pixel at the $k^{th}$ time step is given by
\begin{equation}
x_k = \left[\begin{array}{c} \Re \{C\{A g_k\}\} \\ \Im\{C\{Ag_k\}\} \end{array}\right].
\end{equation}
Assuming our measurement noise, $n_k$, is additive the observation is given by
\begin{equation}
z_k = H_k x_k + n_k.\label{eq:noisysense}
\end{equation}

The true electric field, $x_k$, is the unknown in \eq{eq:noisysense}. Using the observation matrix, $H_k$, we seek an estimate of the electric field, $\hat x_k$, based on the noisy intensity measurements defined in \eq{eq:z}. The estimator will attempt to minimize the error between the estimated and true state. Since the control system does not have access to the true state, the metric defining this error is necessarily based off the measurement residual, $H_k \hat x_k - z_k$. Given a batch of measurements at the current time step, $k$, the cost function used to define the DM diversity algorithm is given by
\begin{equation}
J = \frac{1}{2} [H_k\hat x_k - z_k]^T R_k^{-1}  [H_k\hat x_k - z_k], \label{eq:batchcost}
\end{equation}
where we have weighted the cost function by the measurement covariance, defined as
\begin{equation}
R_k = E[n_k n_k^T].\label{eq:rdef}
\end{equation}
The solution that minimizes the weighted quadratic form of the estimated measurement residual is given by the left pseudo-inverse of $H_k$,
\begin{equation}
\hat x_k = (H_k^TR_k^{-1}H_k)^{-1}H_k^TR_k^{-1} z_k.\label{eq:weightbatch}
\end{equation}
If the measurement noise is assumed to be white, gaussian noise with a standard deviation of $\sigma_k$, the measurement covariance is given by
\begin{equation}
R_k = \sigma_k^2 \, \eye,
\end{equation}
and the state estimate provided by calculating the left-pseudo inverse reduces to
\begin{equation}
\hat x_k = (H_k^TH_k)^{-1}H_k^T z_k.\label{eq:batch}
\end{equation}
With an estimator in hand we wish to quantify its uncertainty by computing its covariance, defined as the expected value of the square estimate error, $x_k - \hat x_k$. Looking back to the weighted form of the estimator in \eq{eq:weightbatch}, the covariance of the estimate is given by
\begin{align}
P_k &= E\left[(x_k-\hat x_k)(x_k-\hat x_k)^T\right] \label{eq:pdef}\\
&= \left(H_k^T R_k^{-1} H_k\right)^{-1}. \label{eq:batchcov}
\end{align}
Since this estimator relies on a new batch of measurements at each control step, the covariance also resets after every control step. Thus the uncertainty in the estimate is tied to the SNR and the quality of the observation (the meaning of which is discussed further in \S\ref{sec:optprobes}) in that particular set of measurements.

With this formulation of the estimator, we can estimate the real and imaginary parts of the aberrated field at each pixel in the image plane with least-squares minimal error on the measurement residual, and by extension minimize the error between the estimated and true state of the image plane electric field. A minimum of 2 image pairs must be used to produce a unique estimate that minimizes the cost function in \eq{eq:batchcost}. The probe shapes rely on an analytical form intended to evenly modulate two rectangular regions of width $w_x$ and height $w_y$, offset from the optical axis by $(a,b)$ \cite{giveon2007broadband,groff_thesis}. Appealing to the Fourier convolution theorem, the DM perturbation that will modulate the image plane in this manner is given by
\begin{equation}
\phi(u,v) = \sinc(w_xu)\sinc(w_yv)\cos(au + \theta_u)\cos(bv + \theta_v),\label{eq:probe}
\end{equation}
where $\theta_u$ and $\theta_v$ are arbitrary phase shifts that are changed for each probe pair. Practically, we find that 4 probe pairs must be used to get a good enough estimate at the Princeton HCIL. This is partially a result of the ad-hoc nature with which the phase shift of each probe pair may be computed, and each phase shift is not guaranteed to adequately modulate the aberrated field. Consequently, 8 images are taken per iteration to estimate the electric field when using the DM diversity algorithm. An optimal probing scheme that solves this limitation is discussed in \S\ref{sec:optprobes}.
\begin{figure}[ht!]
\centering
\subfigure[]{\includegraphics[width = 0.3\textwidth,clip=true,trim=.05in 0in .35in 0in]{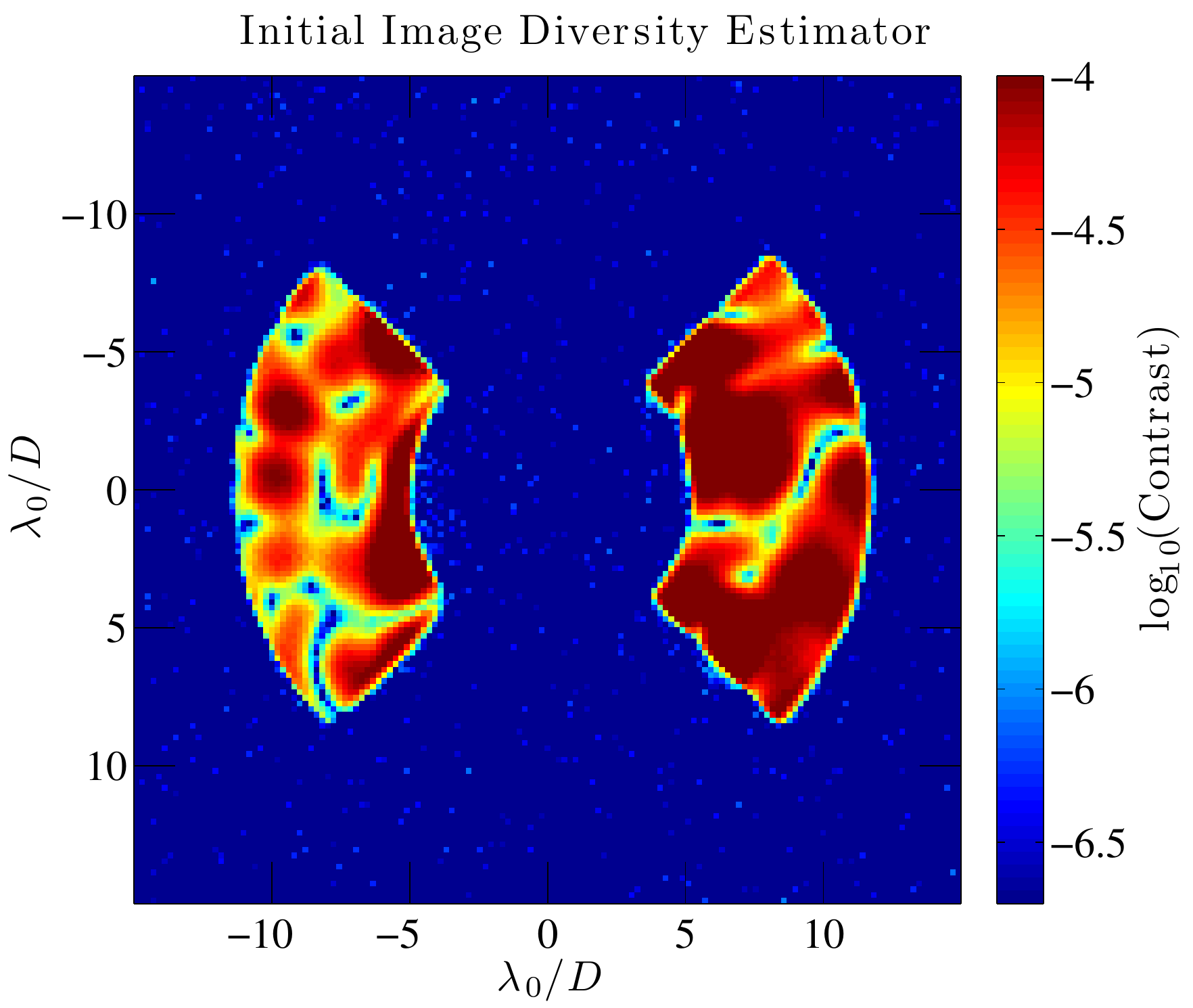}\label{mono_initial}}
\subfigure[]{\includegraphics[width = 0.32\textwidth]{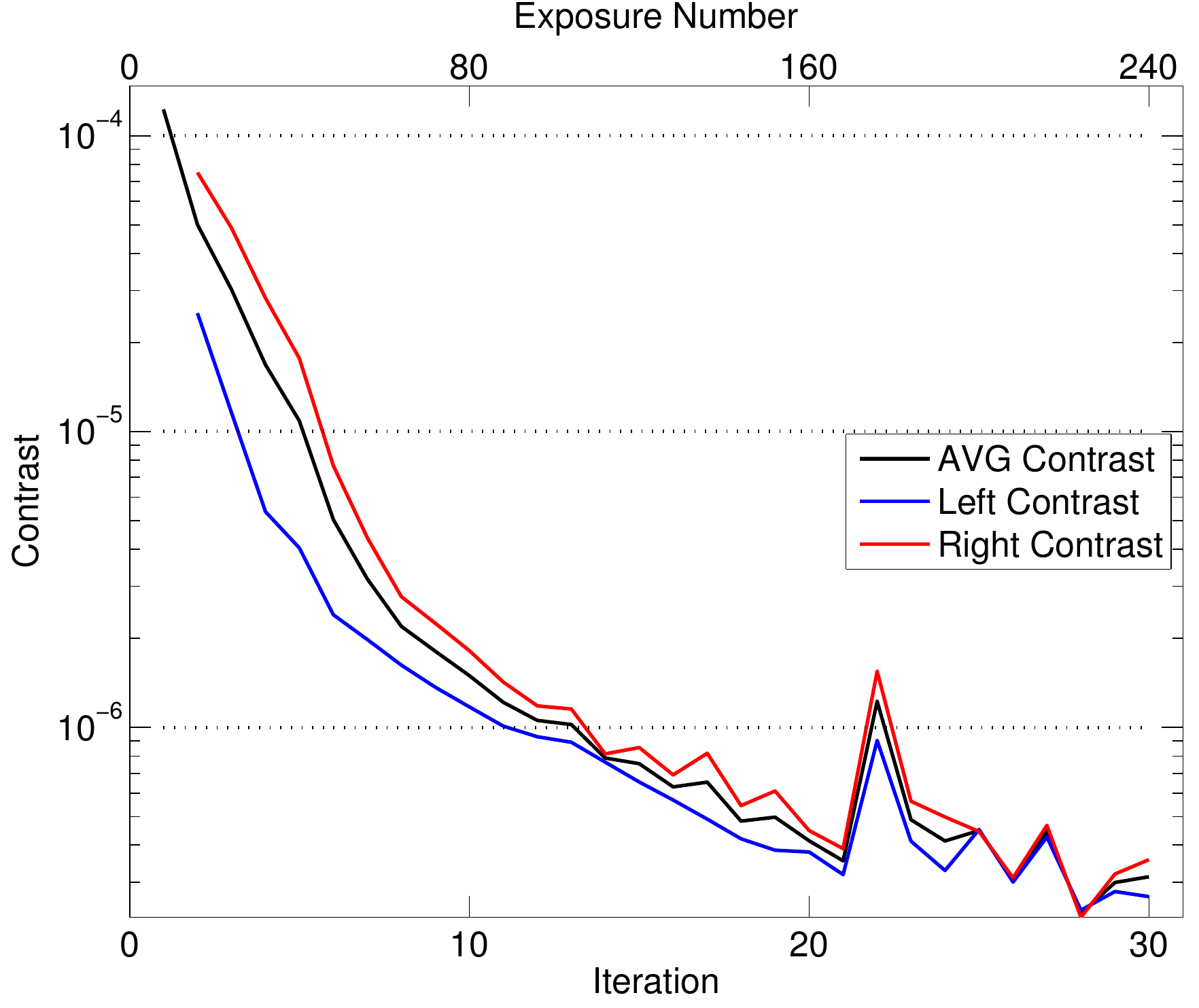}\label{mono_contrast}}
\subfigure[]{\includegraphics[width = 0.3\textwidth,clip=true,trim=.05in 0in .35in 0in]{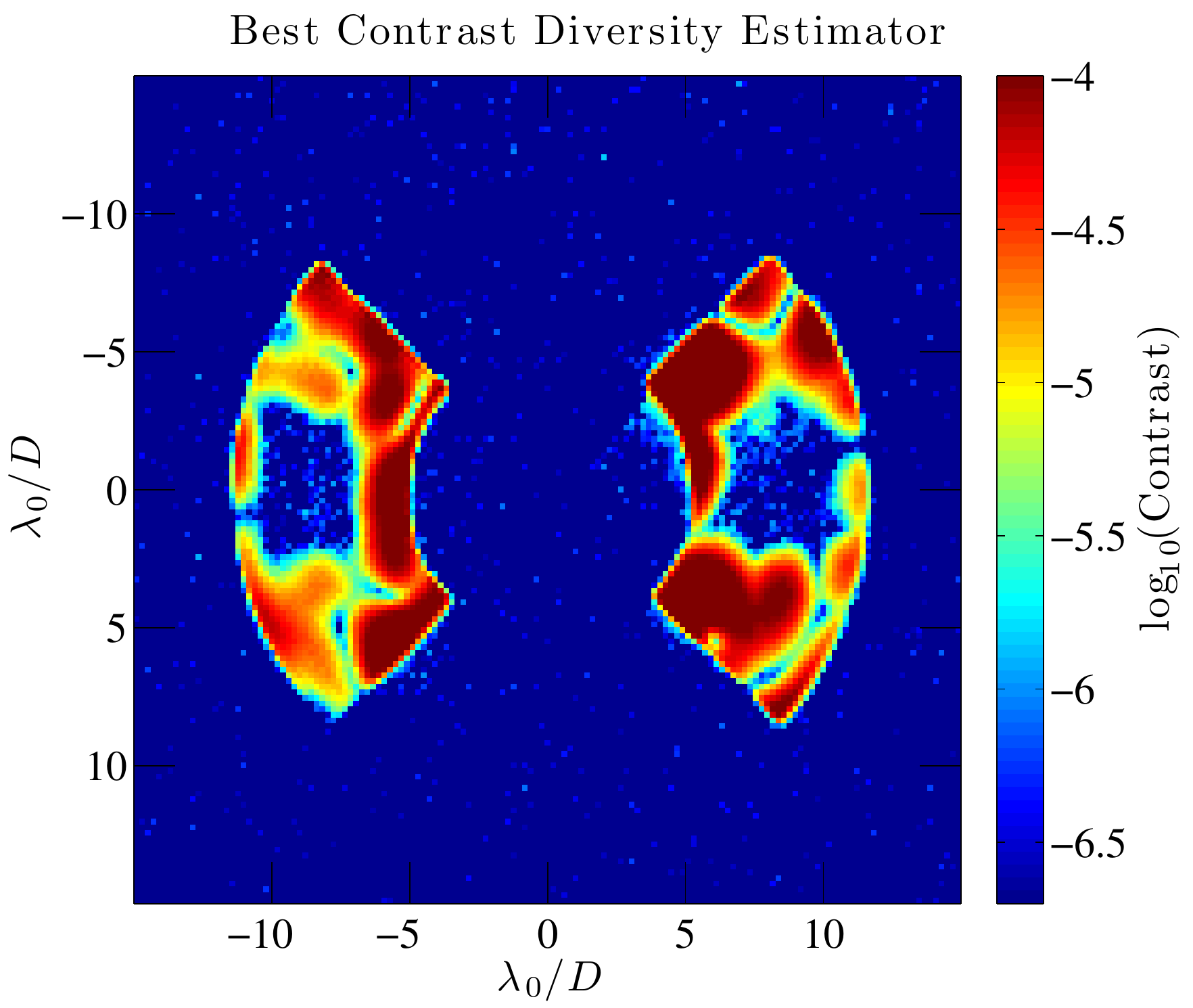}\label{mono_corrected}}
\caption{Experimental results of sequential DM correction using the DM diversity estimation algorithm.  The dark hole is a square opening from 7--10 x -2--2 $\lambda/D$ on both sides of the image plane.  (a) The aberrated image.  (b)  Contrast plot. (c) The corrected image. Image units are log(contrast). }\label{diversity}
\end{figure}

Experimental results using the DM diversity estimator with the stroke minimization algorithm are shown in \fig{diversity}. The laboratory starts at an initial contrast of $1.23\times10^{-4}$, \fig{mono_initial}. Using the least-squares estimation technique it is capable of reaching an average contrast of $2.3\times10^{-7}$ in a (7-10)x(-2-2) $\lambda/D$ region within 30 iterations (\fig{mono_corrected}) on both sides of the image plane, a unique capability that is a result of the two deformable mirrors in the system. In 20 iterations of the algorithm, requiring a total of 160 estimation exposures, the system reached a contrast level of $3.5 \times 10^{-7}$.

\section{Constructing the Kalman Filter Estimator}\label{sec:filter}
\begin{figure}[ht]
\centering
\includegraphics[width = 0.6\textwidth]{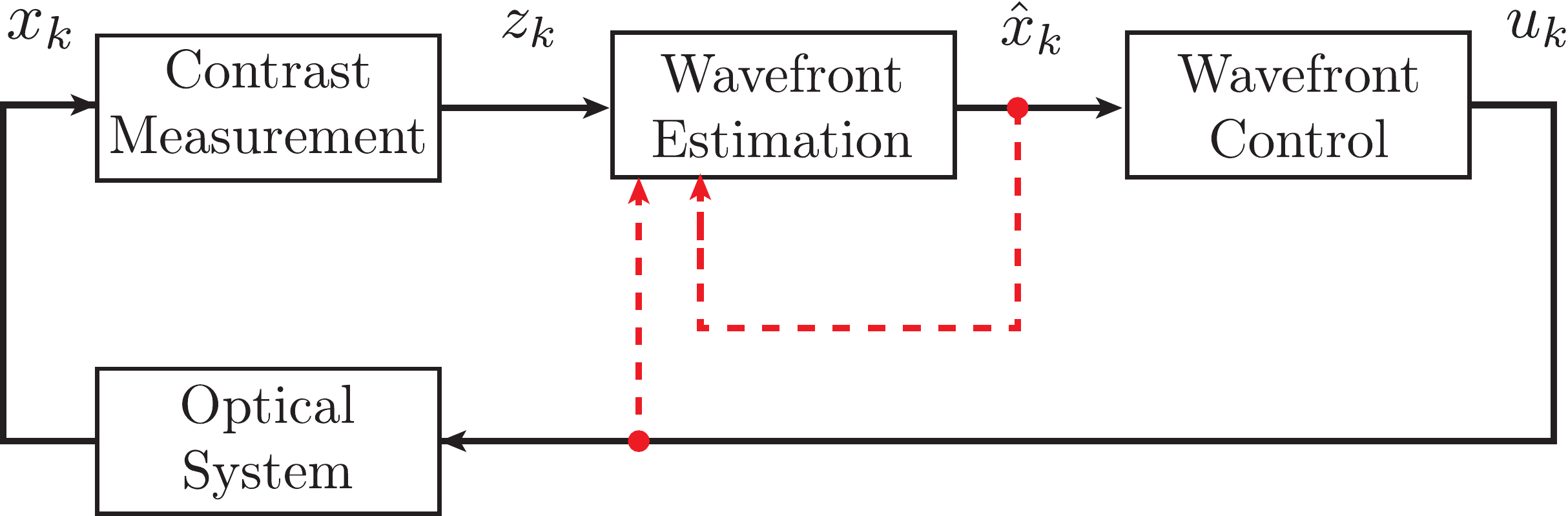}
\caption[Feedback Block Diagram]{Block diagram of a standard FPWC control loop. At time step $k$, only the intensity measurements, $z_k$, provide any feedback to estimate the current state, $x_k$. The dashed lines show additional feedback included in this paper from the prior electric field (or state) estimate, $\hat x_k$, and the control signal, $u_k$. }\label{fig:feedback}
\end{figure}

The DM diversity algorithm \cite{giveon2011pair} is quite effective, but is limited in that it does not close the loop on the state estimate, as shown in \fig{fig:feedback}. Therefore all state estimate information, $\hat x$, acquired about the electric field in the prior control step is lost, requiring a full set of probe images to estimate the field at each time step. 
In addition to being very costly with regard to exposures, the measurements will become progressively noisier as higher contrast levels are reached. Including feedback of the state estimate at each iteration will add a degree of robustness to new, noisy measurements and will reduce the number of exposures required to update the electric field estimate at each control step. Thus, incorporating prior history means the estimator will reach its final contrast target with a minimal set of exposures and will be less sensitive to poor SNR measurements as the controller suppresses the field. To include state feedback we must extrapolate the controller's perturbation to field from the last control step using the model in \S\ref{HCILmodel}. Since the estimate will now be formed by combining an extrapolation of the prior estimate with a smaller set of measurement updates, we must consider the relative effect of process and detector noise to optimally combine the two. This is exactly the problem a discrete time Kalman filter solves. The noisy measurement updates, $z_k = y_k + n_k$, will still be difference images of probe pairs with a covariance defined by \eq{eq:rdef}. The conjugate pairs allow us to construct a linear observation matrix, $H_k$. If we were not in a low aberration regime our observer would have to be nonlinear. This is not impossible for a Kalman filter, but can make it highly biased \cite{stengel1994optimal} and computationally expensive.

Like the DM diversity estimator \cite{giveon2007broadband,giveon2011pair}, the Kalman filter produces an estimate with least-squares minimal error. Since the Kalman filter operates on the estimate in closed loop, the weighted cost function given in \eq{eq:batchcost} will not adequately represent the error contributions in the system. To account for the noise included in forward propagating our prior state estimate history in time, we must also include an estimate of the state covariance, defined in \eq{eq:pdef}. Since the covariance as yet has not been updated with a measurement, we denote it as $P_k(-)$, the minus sign indicating that it is an extrapolation from our model. Defining the error as both the difference between the noisy observation and the estimated observation, $H_k\hat x_k (+) -z $, and the difference between our current estimate and the estimate extrapolation, $\hat x_k -  \hat x_k(-)$, the new quadratic cost function is given by
\begin{equation}
J = \frac{1}{2} \left[\hat x_k - \hat x_k(-)]^T P_k(-)^{-1} [\hat x_k - \hat x_k(-)\right] + \frac{1}{2} \left[H_k\hat x_k - z_k\right]^T R_k^{-1}  \left[H_k\hat x_k - z_k\right]. \label{eq:kalmancost}
\end{equation}
We formulate the cost in matrix form as
\begin{align}
J &= \frac{1}{2}\begin{bmatrix} \hat x_k - \hat x_k(-) \\ H_k \hat x_k - z_k\end{bmatrix}^T \begin{bmatrix} P_k(-) & 0 \\ 0 & R_k \end{bmatrix}^{-1} \begin{bmatrix} \hat x_k - \hat x_k(-) \\ H_k \hat x_k - z_k\end{bmatrix} \\
&= \frac{1}{2} (\tilde H_k \hat x_k - \tilde z_k)^T \tilde R_k^{-1}  (\tilde H_k\hat x_k - \tilde z_k), \label{eq:kalmanquad}
\end{align}
where we have now defined a new set of augmented matrices as
\begin{align}
\tilde H_k &= \begin{bmatrix}\eye\\H_k\end{bmatrix}, \label{eq:augH} \\
\tilde z_k &= \begin{bmatrix}\hat x_k(-) \\ z_k \end{bmatrix},\\
\tilde R_k &= \begin{bmatrix} P_k(-) & 0 \\ 0 & R_k \end{bmatrix}. \label{eq:augR}
\end{align}
Evaluating the partial derivative of \eq{eq:kalmanquad} the state estimate update is given by
\begin{equation}
\hat x_k(+)=  \hat x_k(-) + P_k(-) H_k^T\left[H_kP_k(-)H_k^T + R_k\right]^{-1} \left[z_k - H_k \hat x_k(-) \right]. \label{eq:deriveupdate}
\end{equation}
From \eq{eq:deriveupdate}, we define the optimal gain to be
\begin{equation}
K_k = P_k(-) H_k^T[H_kP_k(-)H_k^T + R_k]^{-1}.\label{eq:gainderive}
\end{equation}
Finally, we update the state covariance estimate, $P_k(+)$, by applying \eq{eq:deriveupdate} to the expected value of the covariance,
\begin{equation}
P_k(+) = E[(\hat x_k(+) - x_k)(\hat x_k(+) - x_k)^T],
\end{equation}
which gives
\begin{equation}
P_k(+) = [P_k(-)^{-1} + H_k^T R_k^{-1} H_k ]^{-1}.\label{eq:covderive}
\end{equation}

To complete the filter we need to propagate the prior estimate, $\hat x_{k-1}(+)$, to the current time step. The filter extrapolates to the current state estimate, $\hat x_k(-)$, by applying a time update to the prior state estimate via the state transition matrix, $\Phi_{k-1}$, and numerically propagating the control output from stroke minimization at the prior iteration, $u_{k-1}$,  via a linear transformation described by $\Gamma_{k-1}$. We also have a disturbance from the process noise, $w_{k-1}$, which is propagated to the current state of the electric field via the linear transformation, $\Lambda_{k-1}$. Assuming these components are additive, the state estimate extrapolation is given by
\begin{equation}
\hat x_k (-) = \Phi_{k-1} \hat x_{k-1}(+) + \Gamma_{k-1} u_{k-1} + \Lambda_{k-1} w_{k-1}. \label{eq:estpropproc}
\end{equation}
To determine these matrices, we apply the linearized optical model developed in \S\ref{HCILmodel}. Using a linearized model avoids generating arbitrary bias in the estimate at each pixel, a common problem with a nonlinear filter \cite{gelb1974optimal}. The first step in propagating the state forward in time is to update any dynamic variation between the discrete time steps with the state transition matrix, $\Phi_{k-1}$. In this system, $\Phi_{k-1}$ captures any variation of the field due to temperature fluctuations, vibration, or air turbulence that perturb the optical system. To simplify the model, we recognize that there is no reliable way to measure or approximate small changes in the optical system over time with alternate sensors; we assume that the state remains constant between control steps, making the state transition matrix the identity, $\Phi_{k-1} = \Phi = \mathcal{I}$. However, it is very important to note that $\Phi_{k-1}$ provides a simple mechanism to incorporate time updates from alternative high speed sensors, such as residuals measured by the adaptive optics wavefront sensor, between image plane measurements. Each submatrix for $\Gamma$, shown in Table~\ref{filtertable}, is of dimension $2 \times 2N_{DM}$. Every pair of rows is found by separating the real and imaginary parts of \eq{eq:g} into individual rows. Making the standard assumption that the process noise is gaussian white noise, the expected value of the state when we extrapolate is
 \begin{equation}
 \hat x_k (-) = \Phi_{k-1} \hat x_{k-1}(+) + \Gamma_{k-1} u_{k-1}. \label{eq:estprop}
\end{equation}
It's associated covariance extrapolation is then given by
\begin{equation}
P_k(-) = \Phi_{k-1} P_{k-1}(+) \Phi_{k-1}^T + Q_{k-1}. \label{eq:covprop}
\end{equation}

Combining \eq{eq:deriveupdate}, \eq{eq:gainderive}, and, \eq{eq:covderive} with the extrapolation equations, this form of the filter consists of five equations that describe the state estimate extrapolation, covariance estimate extrapolation, filter gain computation, state estimate update, and covariance estimate update at the $k^{th}$ iteration \cite{stengel1994optimal}:
\begin{align}
\hat x_k (-) &= \Phi_{k-1} \hat x_{k-1}(+) + \Gamma_{k-1} u_{k-1}. \label{eq:estextrap}\\
P_k(-) &= \Phi_{k-1} P_{k-1}(+) \Phi_{k-1}^T + Q_{k-1} \label{eq:covextrap}\\
K_k &= P_k(-)H_k^T \left[H_kP_k(-)H_k^T + R_k \right]^{-1}\label{eq:gain}\\
\hat x_k(+) &= \hat x_k(-) + K_k \left[z_k - H_k \hat x_k(-) \right] \label{eq:estupdate}\\
P_k(+) &= \left[P_k(-)^{-1} + H_k^T R_k^{-1}H_k \right]^{-1}\label{eq:covupdate}
\end{align}

The focal plane measurements $z_k$ are identical to that of \S\ref{diversity}, and are constructed into a column vector of  $j$ difference images for $n$ pixels. Likewise $H_k$ takes on a similar form, and is a matrix constructed from the effect of a specific deformable mirror shape $\phi_{j,k}$ on the real and imaginary parts of the electric field in the image plane. Finally, we compute the covariance update, $P_k(+)$, based on the added noise from the new measurements. As with the DM diversity estimator, the estimated state is a column vector of the real and imaginary parts of the electric field at each pixel of the dark hole. The control signal $u$ remains a column vector of the DM actuator strengths as defined in \S\ref{HCILmodel}, with DM1 commands being stacked on top of the DM2 commands. The dimension and form of the rest of the filter equations follows from these. They are all listed in detail in Table~\ref{filtertable} and Table~\ref{noisetable}. Since we are only considering process noise at the DMs, the process disturbance $w$ is a vertical stack of the variance expected from each actuator. The only remaining step critical to the filter's performance is initializing the covariance, $P_0$. In our system this cannot be measured, so we must initialize with a reasonable guess. 

The Kalman filter gain, \eq{eq:gain}, optimally combines the prior estimate history with measurement updates to minimize the total error contributions based on the expected state and measurement covariance. A fundamental property of the Kalman filter is that the optimal gain, \eq{eq:gain}, is not based on measurements, but rather estimates of the state covariance, $P_k(-)$, process noise from the actuation $Q_{k-1}$, and sensor noise $R_k$. This means that the optimality of the estimate is closely related to the accuracy and form of these matrices; this will be discussed in \S\ref{sec:noise}. Much like the batch process method the purpose of the gain in the Kalman filter is to minimize a quadratic cost function, \eq{eq:kalmancost}, but it is also subject to the constraining dynamic equations defining $\hat x_k(-)$ and $P_k(-)$. However, looking at \eq{eq:augH} there is a major advantage of the Kalman filter in its minimization of the cost function. For the batch process method, the observation matrix will be underdetermined if less than 2 probe pairs are applied. Thus the solution to cost function in \eq{eq:batchcost} is non-unique. However the solution for the Kalman filter is guaranteed to be critically determined even with no measurement update, and $\tilde H_k$ only requires a single measurement to provide an overdetermined solution to the cost function in \eq{eq:kalmanquad}. Thus, the Kalman filter is formulated in such a way that it solves a least squares error, left pseudo-inverse problem, regardless of the number of measurements taken. This gives us the freedom to minimize the total number of exposures required to estimate the field with enough accuracy to reach the target contrast level.

Referring back to \eq{eq:batchcov}, we see that in both cases the covariance depends on the current observation, but the Kalman filter is also a function of the prior state covariance. Looking at \eq{eq:covupdate} the final term,  $H_k^T R_k^{-1} H_k$, is guaranteed to be positive definite, meaning that additional measurements can do nothing but reduce the magnitude of the covariance, $P_k(+)$. In the event of a measurement with a poor SNR the covariance may not get better, but it is guaranteed not to get worse because $H_k^TR_k^{-1}H_k$ is positive definite. This is indicative of the estimator's robustness to bad measurements at later time steps. In contrast, \eq{eq:batchcov} demonstrates that if the left-pseudo inverse solution is given measurements with a poor SNR the new estimate will have a large covariance at this time step. In this case the control will not be effective, which is why we often see jumps in contrast when the DM Diversity estimator reaches low contrast levels, as seen in \fig{diversity}. In the case of the Kalman filter, the high covariance contributed by a poor measurement update is dampened by the contribution of prior covariance estimates via $P_k(-)$, stabilizing the state estimate and its covariance in the event of a bad measurement. Since we cannot guarantee that a probe will provide a high enough SNR, particularly at high contrast levels where we have reduced the number of photons per speckle, this is an extremely attractive component of the Kalman filter estimator. 
 \begin{table}[ht!]
\centering
   \begin{tabular}{ cc } 
         \toprule
      Matrix & Dimension \\
      \midrule
      	$\Phi = \mathcal I $& $(2 \cdot N_{pixels}) \times (2 \cdot N_{pixels})$\bigskip\\
	$\Gamma  = \begin{bmatrix} 
	\begin{bmatrix} \Re \{G_{DM1}\} & \Re \{G_{DM2}\} \\  \Im \{G_{DM1}\} & \Im \{G_{DM2}\} \end{bmatrix}_1 \\
	\vdots \\
	\begin{bmatrix} \Re \{G_{DM1}\} & \Re \{G_{DM2}\} \\  \Im \{G_{DM1}\} & \Im \{G_{DM2}\} \end{bmatrix}_n
	\end{bmatrix}$ &  $(2 \cdot N_{pixels}) \times(2 \cdot N_{DM})$\bigskip\\
	$H_k = diag \left(\begin{bmatrix} \Re \{G_{DM2} \phi_{1,k}\} & \Im \{G_{DM2} \phi_{1,k}\} \\ \vdots & \vdots \\ \Re \{G_{DM2} \phi_{j,k}\} & \Im \{G_{DM2} \phi_{j,k}\} \end{bmatrix}_{n} \right) $ &  $(N_{pixels})\cdot(N_{pairs}) \times(2 \cdot N_{pixels}) $\bigskip\\
	$K_k$ is computed & $(2 \cdot N_{pixels})  \times(N_{pixels})\cdot(N_{pairs})$\bigskip\\
	$u = \begin{bmatrix} DM1 \\ DM2 \end{bmatrix}$ & $ (2 \cdot N_{DM}) \times 1$\bigskip\\
	$z = \begin{bmatrix} 
		\begin{bmatrix} I_1^+ - I_1^- \\ \vdots \\ I_j^+ - I_j^- \end{bmatrix}_{1} \\ 
		\vdots \\  
		\begin{bmatrix} I_1^+ - I_1^- \\ \vdots \\ I_j^+ - I_j^- \end{bmatrix}_{n} 
		\end{bmatrix}$ & $(N_{pixels})\cdot (N_{pairs}) \times 1$\bigskip\\
   	$\hat x^T = \begin{bmatrix} 
			\begin{bmatrix} \Re \{E_1\} & \Im\{E_1\} \end{bmatrix} &
			\cdots & 
			\begin{bmatrix} \Re \{E_n\} & \Im\{E_n\} \end{bmatrix}
			\end{bmatrix}^T$  & $(2 \cdot N_{pixels}) \times 1$ \bigskip\\
  \bottomrule
   \end{tabular}
   \caption{Definition of all propagation Matrices.  $N_{DM}$ is the number of actuators on a single DM, $N_{pixels}$ is the number of pixels in the area targeted for dark hole generation, and $N_{pairs}$ is the number of image pairs taken while applying positive and negative shapes to the deformable mirror. $j$ indexes the probe shape, $k$ the discrete time step, and $n$ the pixel the dark hole.}
   \label{filtertable}
\end{table}

 \begin{table}[ht!]
\centering
   \begin{tabular}{ cc } 
         \toprule
      Matrix & Dimension \\
      \midrule
      	$P_0 = E[(x_0 - \hat x_0)(x_0 - \hat x_0)^T]$ & $(2 \cdot N_{pixels}) \times(2 \cdot N_{pixels})$\bigskip\\
	$Q_k = \Lambda E[w_kw_k^T] \Lambda^T$ & $(2 \cdot N_{pixels}) \times(2 \cdot N_{pixels})$\bigskip\\
	$R_k = E[n_k n_k^T]$ & $(N_{pixels})\cdot (N_{pairs}) \times (N_{pixels})\cdot(N_{pairs}) $ \bigskip\\
	$\Lambda = \Gamma$ & $(2 \cdot N_{pixels}) \times(2 \cdot N_{DM})$\bigskip\\
	$w = \begin{bmatrix} \sigma_{DM1} \\ \sigma_{DM2} \end{bmatrix}$ & $(2 \cdot N_{DM}) \times 1$\bigskip\\
  \bottomrule
   \end{tabular}
   \caption{Definition of all noise Matrices.  $N_{DM}$ is the number of actuators on a single DM, $N_{pixels}$ is the number of pixels in the area targeted for dark hole generation, and $N_{pairs}$ is the number of image pairs taken while applying positive and negative shapes to the deformable mirror. $j$ indexes the probe shape, $k$ the discrete time step, and $n$ the pixel the dark hole.}
   \label{noisetable}
\end{table}

\section{Iterative Kalman Filter}\label{sec:iterative}
Since we had to linearize the field to obtain $\Gamma_{k-1}$ and $H_k$, we iterate the filter on itself to account for some of the nonlinearity not captured in the initial linearization. We do so by feeding the newly computed state $\hat x_k(+)$ and covariance update $P_k(+)$ back into the filter again, setting $u_{k-1}$ to zero. Applying feedback to the filter in this manner is purely computational, and comes at no cost with regard to exposures. For sufficiently small control this additional computation will account for nonlinearity in the actuation and better filter noise in the system, limited only by the accuracy of the observation matrix, $H_k$. Improving the accuracy at each time step means the correction algorithm will require fewer iterations, and hence fewer exposures, to reach its final contrast target. For a severely photon-limited exoplanet imager, such a reduction minimizes the time required to suppress the speckle field. Following a notation similar to Gelb \cite{gelb1974optimal}, the $p^{th}$ iteration (for $p\geq 2$) of numerical feedback into the iterative Kalman filter at the $k^{th}$ control step is
\begin{align}
\hat x_{p,k} (-) &=  \hat x_{p-1,k}(+) \label{eq:iterstate}\\
P_{p,k}(-) &= \Phi_{p-1,k} P_{p-1,k}(+) \Phi_{p-1,k}^T + Q_{p-1,k} \label{eq:itercov}\\
K_{p,k} &= P_{p,k}(-)H_{p,k}^T \left[H_{p,k}P_{p,k}(-)H_{p,k}^T + R_{p,k} \right]^{-1}\label{eq:itergain}\\
\hat x_{p,k}(+) &= \hat x_{p,k}(-) + K_{p,k} \left[z_{p,k} - H_{p,k} \hat x_{p,k}(-) \right] \label{eq:iterupdate}\\
P_{p,k}(+) &= \left[P_{p,k}(-)^{-1} + H_{p,k}^T R_{p,k}^{-1}H_{p,k} \right]^{-1}.\label{eq:itercov}
\end{align}

The power of iterating the filter lies in what we are fundamentally trying to achieve. For a successful control signal, we will have suppressed the field. This means that the magnitude of the probe signal will be lower than the control perturbation. This guarantees that $H_k$ will better satisfy the linearity condition than $\Gamma u$. As a result, if we iterate the filter on itself during a given control step we can use the discrepancy between the image predicted by $H_k \hat x_k(+)$ and the measurements, $z_k$, in \eq{eq:iterupdate} to filter out any error due to nonlinear terms not accounted for in $\Gamma$. In this way, we can accommodate a small amount of nonlinearity in our extrapolation of the state without having to resort to a nonlinear, or extended, Kalman filter. This means that in the work presented here we did not have to re-linearize within a single control step, $k$, as would be the case for an iterative extended Kalman filter (IEKF). It also avoided having to concern ourselves with any bias introduced into the estimate by a nonlinear filter. While we have chosen not to in this case, we can move to an IEKF in the future by simply re-linearizing about the current DM shape and iterating on $p$ until the estimate converges. Something very important to point out is that unlike many IEKF problems, this has a unique flexibility with regard to the linearization. A conventional IEKF has been linearized about the state, $\hat x_k$, meaning that the new matrices in the IEKF must remain within an allowable radius of convergence, oftentimes requiring that the estimate be re-linearized at every iteration on $p$ \cite{gelb1974optimal}. Looking back to \eq{eq:linpupnonzero} however, we recall that we have actually linearized about the control signal. Since we did not linearize about the state and the problem assumes no dynamics at this point, we need only re-linearize $\Gamma_{k-1}$ and $H_k$. This re-linearization need not wait for the estimate to begin. Once the control at $(k-1)$ is applied we can re-linearize $\Gamma_{k-1}$ and $H_k$ in parallel with taking new measurements for the current time step, $z_k$.
Thus for an adequately long exposure, re-linearizing these matrices can be scheduled in such a way that they have negligible impact on the time required to estimate the field. Additionally, \eq{eq:covextrap} and \eq{eq:gain} are measurement independent. Thus, their computation can be scheduled in parallel to taking the new measurements, further improving the efficiency of the estimator with regard to time.
\section{Sensor and Process Noise}\label{sec:noise}
Two important design parameters for the performance of the filter are the process noise, $Q_{k-1}$, and the sensor noise, $R_k$. For the laboratory experiments we make reasonable assumptions by appealing to physical scaling of the two largest known sources of error in the system. The sensor noise will be determined by the dark current and read noise inherent to our detector. In a space observatory, and possibly even on the ground, the sensor noise will also be determined by photon noise. We assume that the process noise is dominated by errors in the commanded DM shapes. Thus, the process noise is isolated to poor knowledge of the DM surface, which comes from the inherent nonlinearity in the voltage-to-actuation gain as a function of voltage, the variance in this gain from actuator to actuator, and the accuracy of the superposition model used to construct the mirror surface that covers the 32x32 actuator array of the Boston Micromachines kilo-DM. 
Treating actuation errors as additive process noise, the Kalman filter can account for them in a statistical fashion, rather than deterministically in a physical model. Since there is no physical reasoning to justify varying $Q_k$ at each iteration, it will be kept constant throughout the entire control history ($Q_{k-1} = Q =$ constant). Two versions of process noise can be considered in this case. The first is where there is no correlation between actuators, giving a purely diagonal matrix with a magnitude corresponding to the square of the actuation variance, $\sigma_u$. The second version of $Q$ has symmetric off-diagonal elements, which treats uncertainty due to inter-actuator coupling and errors in the superposition model statistically. As a first step, we will not consider inter-actuator coupling to help avoid a poorly conditioned matrix. This helps guarantee that the Kalman filter itself will be well behaved. Thus the process noise for the filter will be

\begin{equation}
Q = \sigma_{u}^2 \Gamma \mathcal I  \Gamma^T.
\end{equation}

Following Howell \cite{howell2000handbook} the noise from both the incident light and dark noise in a CCD detector follows a Poisson distribution. Since each measurement is a difference of pairwise images the noise is zero-mean and will become more Gaussian as the exposure time increases. We simplify the noise statistics by assuming it is uncorrelated and constant from pixel to pixel.  The measurement error covariance, $R_k$, is thus a diagonal matrix of the mean pixel covariance, $\sigma_{CCD}$, given by
\begin{equation}
R = \frac{\sigma_{CCD}^2}{I_{00}} \mathcal I_{N_{pairs} \times N_{pairs}},\label{eq:sensnoise}
\end{equation}
where $I_{00}$ is the peak count rate of the PSF's core (allowing us to describe $R$ in units of contrast). Having appealed to physical scaling in the HCIL, we now have close approximations of the true process and sensor noise exhibited in the experiment.

\section{Experimental Results for the Kalman Filter Estimator in Monochromatic Light}\label{sec:results}
\begin{figure}[h!]
\centering
\subfigure[]{\includegraphics[width = 0.32\textwidth]{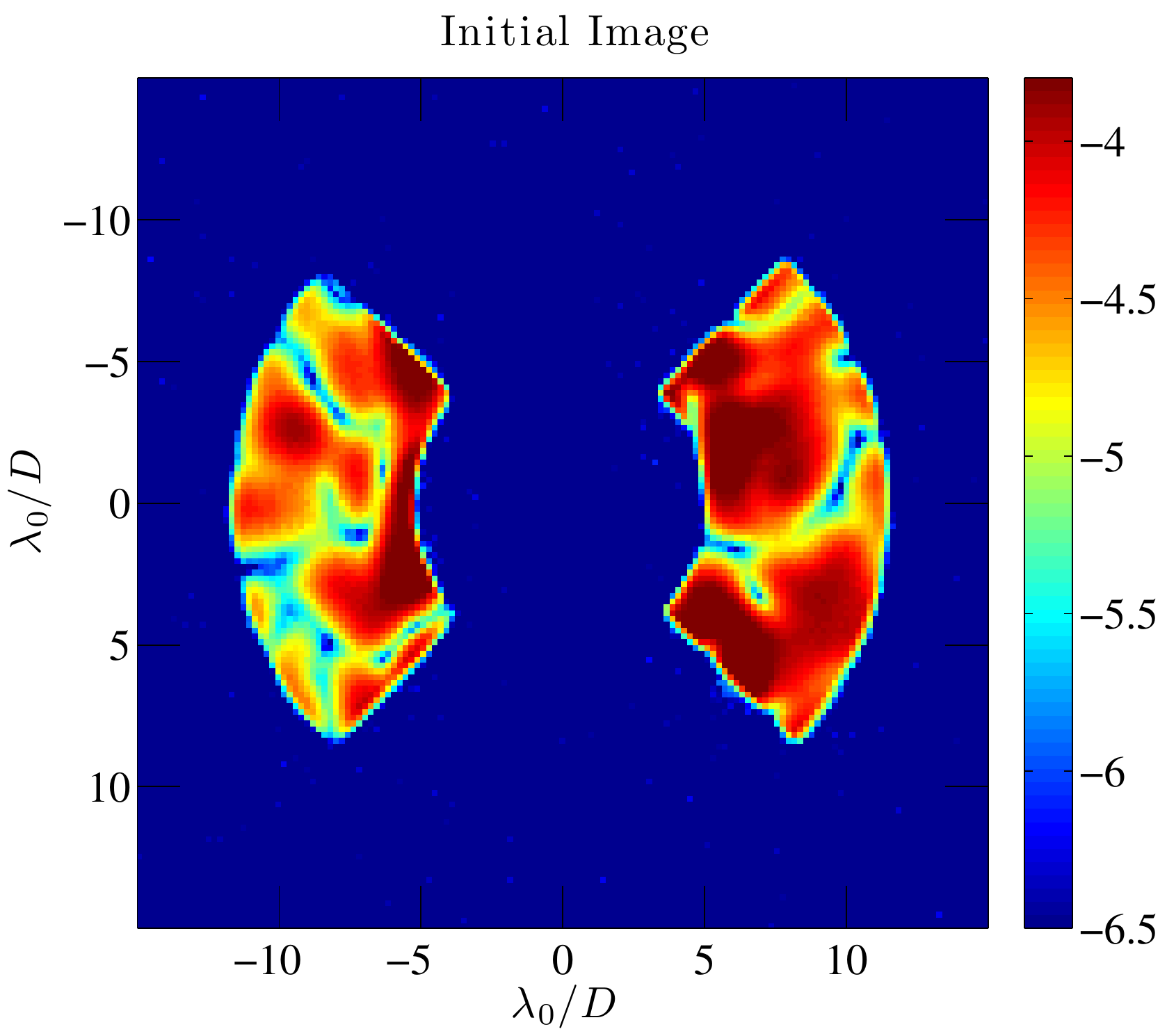}\label{fig:4pairs_initial}}
\subfigure[]{\includegraphics[width = 0.335\textwidth]{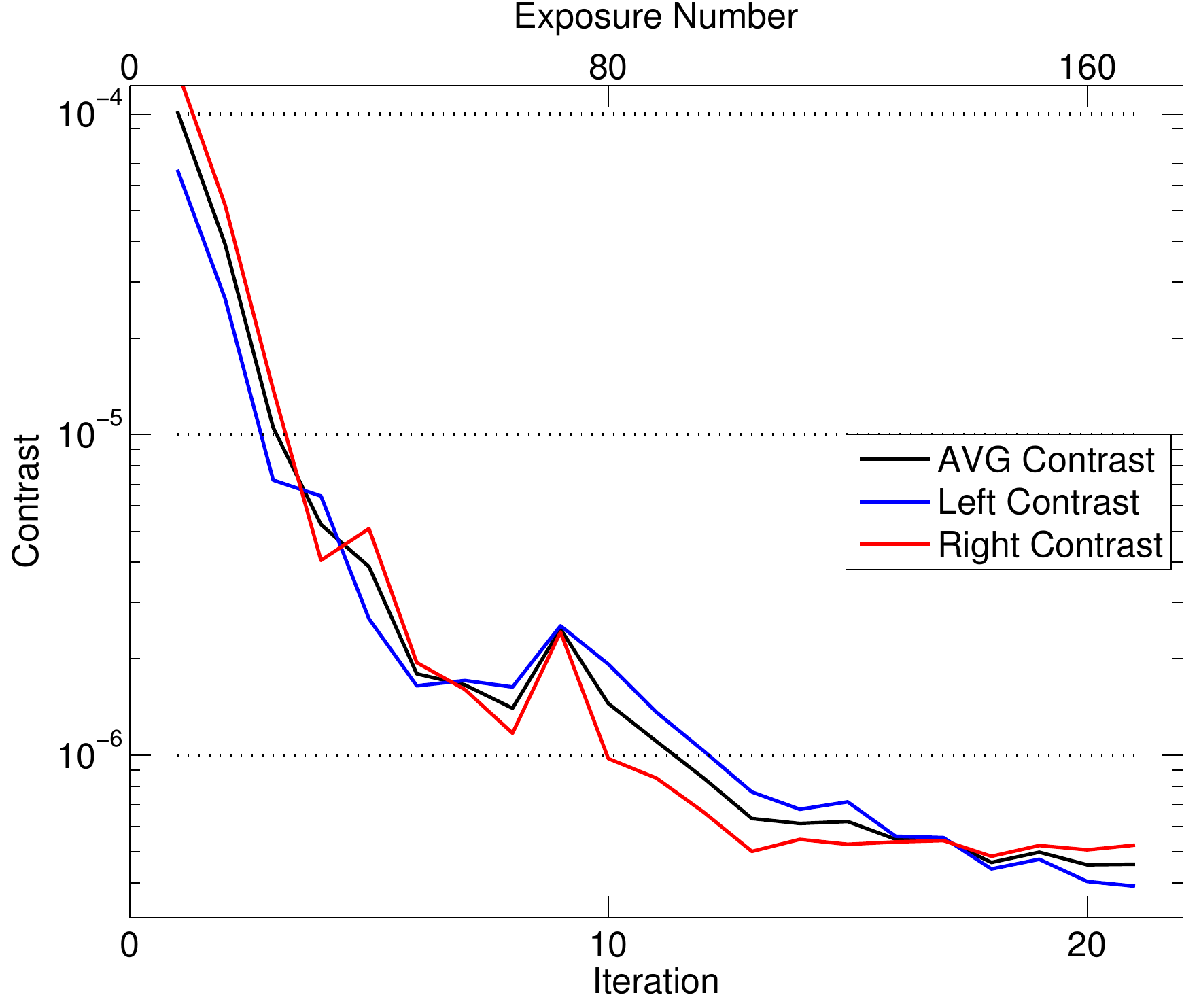}\label{fig:4pairs_contrast}}
\subfigure[]{\includegraphics[width = 0.32\textwidth]{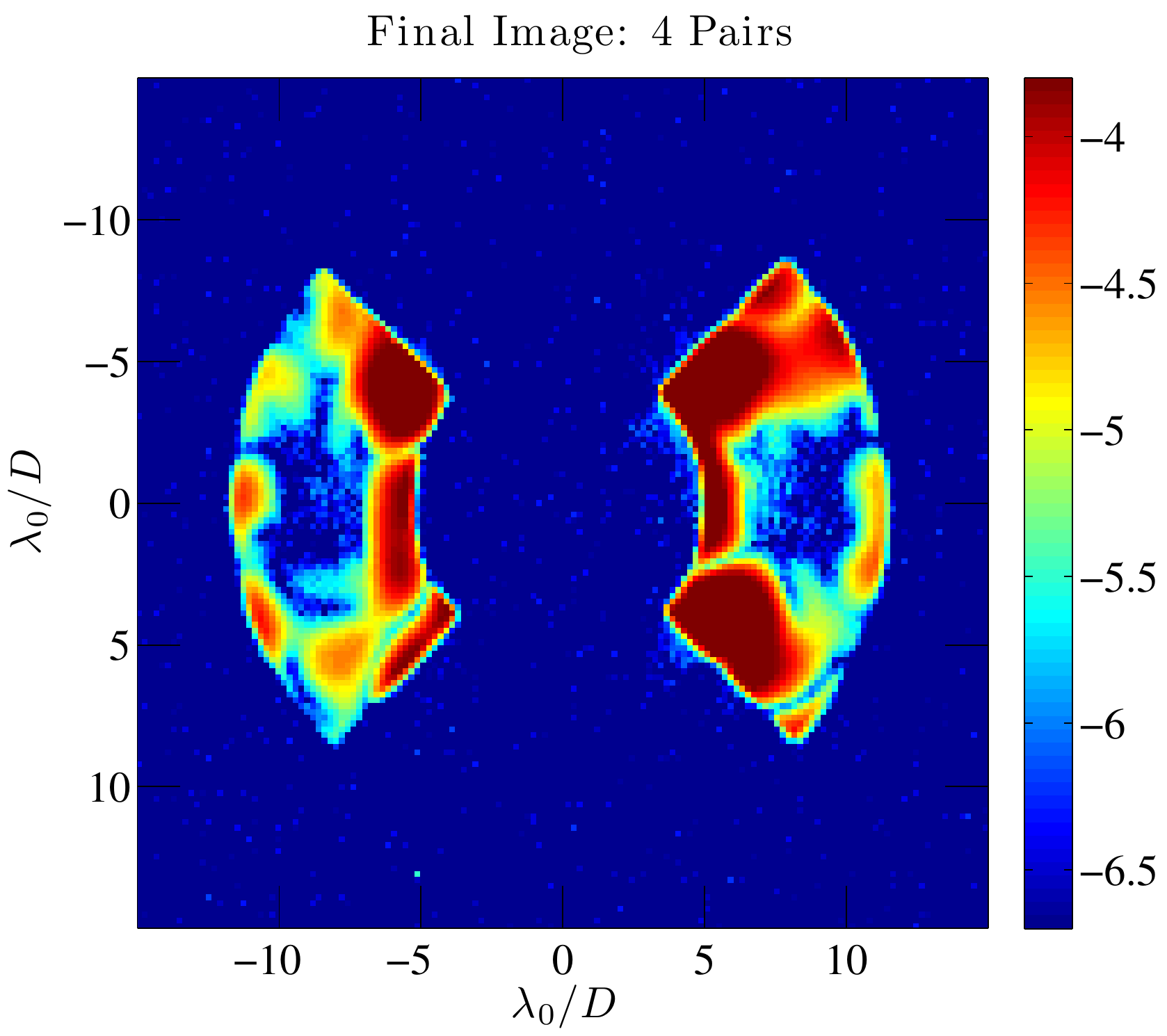}\label{fig:4pairs_final}}
\caption{Experimental results of sequential DM correction using the discrete time extended Kalman filter with 4 image pairs to build the image plane measurement, $z_k$.  The dark hole is a square opening from 7--10 $\times$ -2--2 $\lambda/D$ on both sides of the image plane.  (a) The aberrated image.  (b)  Contrast plot. (c) The corrected image. Image units are log(contrast). }\label{fig:4pairs}
\end{figure}

We corrected the field using the Kalman filter estimator using four, three, two, and one pair of images as a measurement update to assess the degradation in performance as information is lost. We began with four measurements (four image pairs), to compare its performance using the same number of measurements as were used in the DM diversity estimator. Using 4 pairs, the filter achieved a contrast of $4.0 \times 10^{-7}$ in (7-10)x(-2-2) $\lambda/D$ symmetric dark holes within 20 iterations of the controller, shown in \fig{fig:4pairs}. Note that this used a total of $160$ estimation images, which is the same amount of information available to the DM diversity estimator when it achieved a contrast of $3.5 \times 10^{-7}$ in $20$ iterations.

\begin{figure}[h!]
\centering
\subfigure[]{\includegraphics[width = 0.32\textwidth]{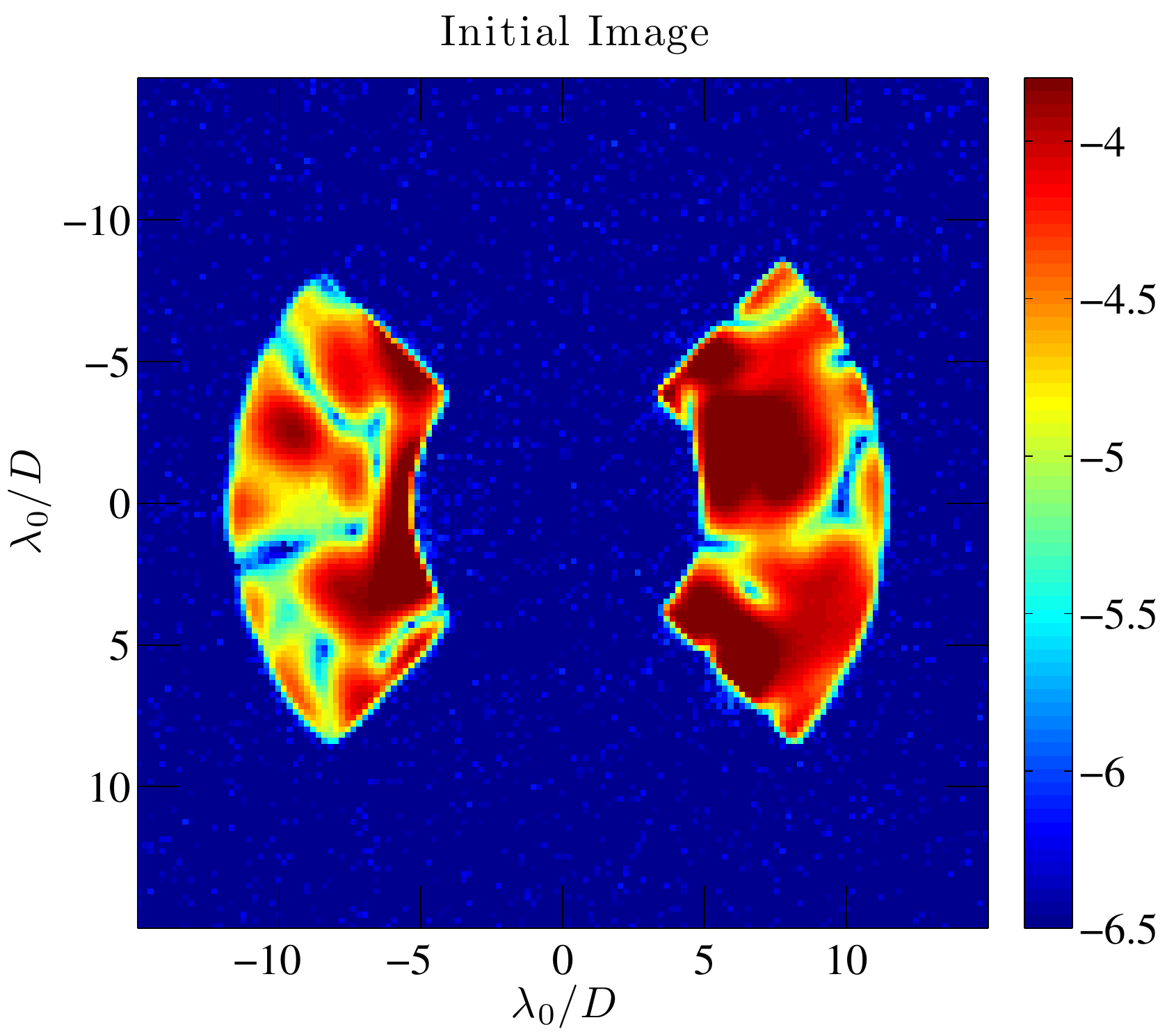}\label{fig:3pairs_initial}}
\subfigure[]{\includegraphics[width = 0.335\textwidth]{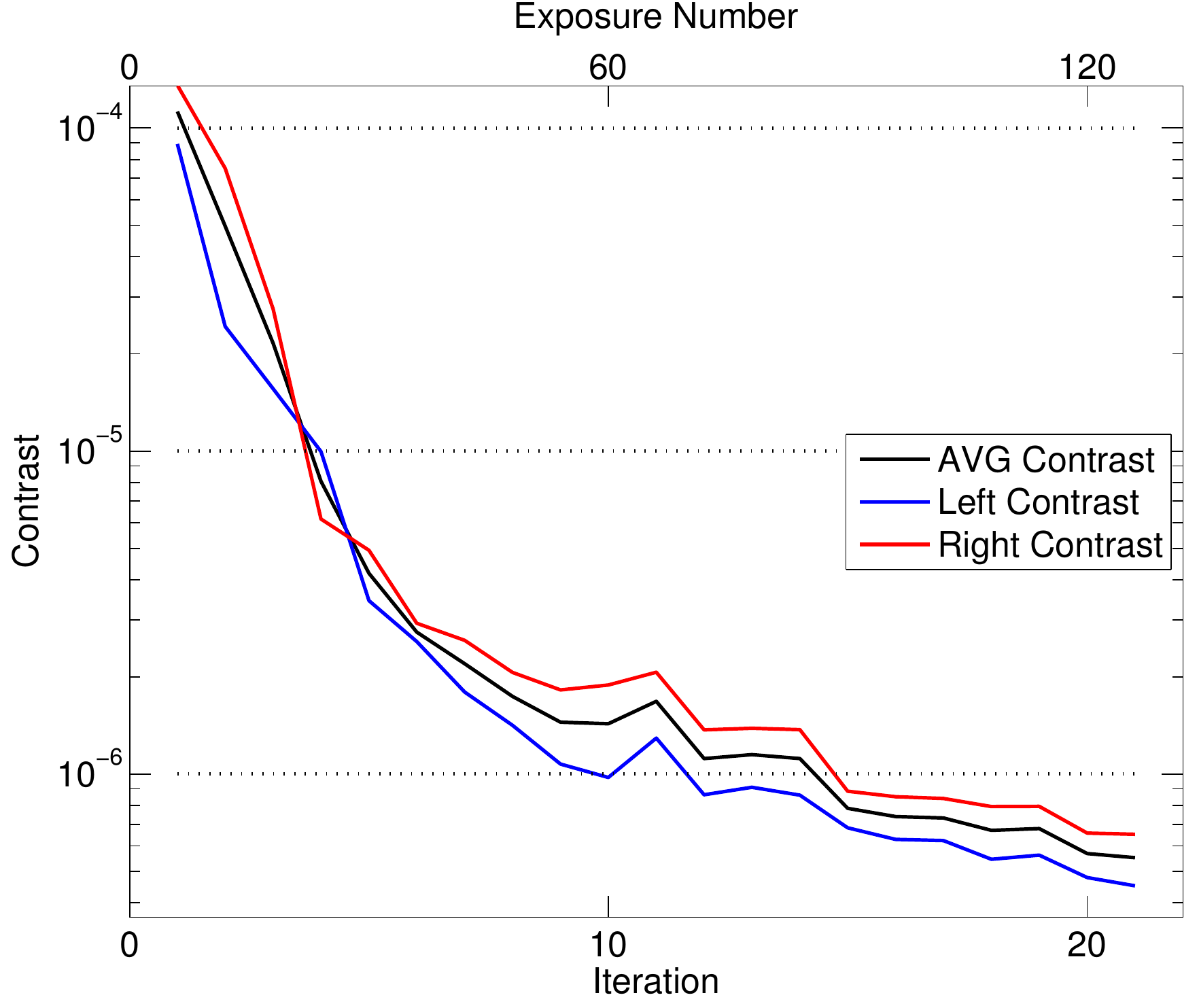}\label{fig:3pairs_contrast}}
\subfigure[]{\includegraphics[width = 0.32\textwidth]{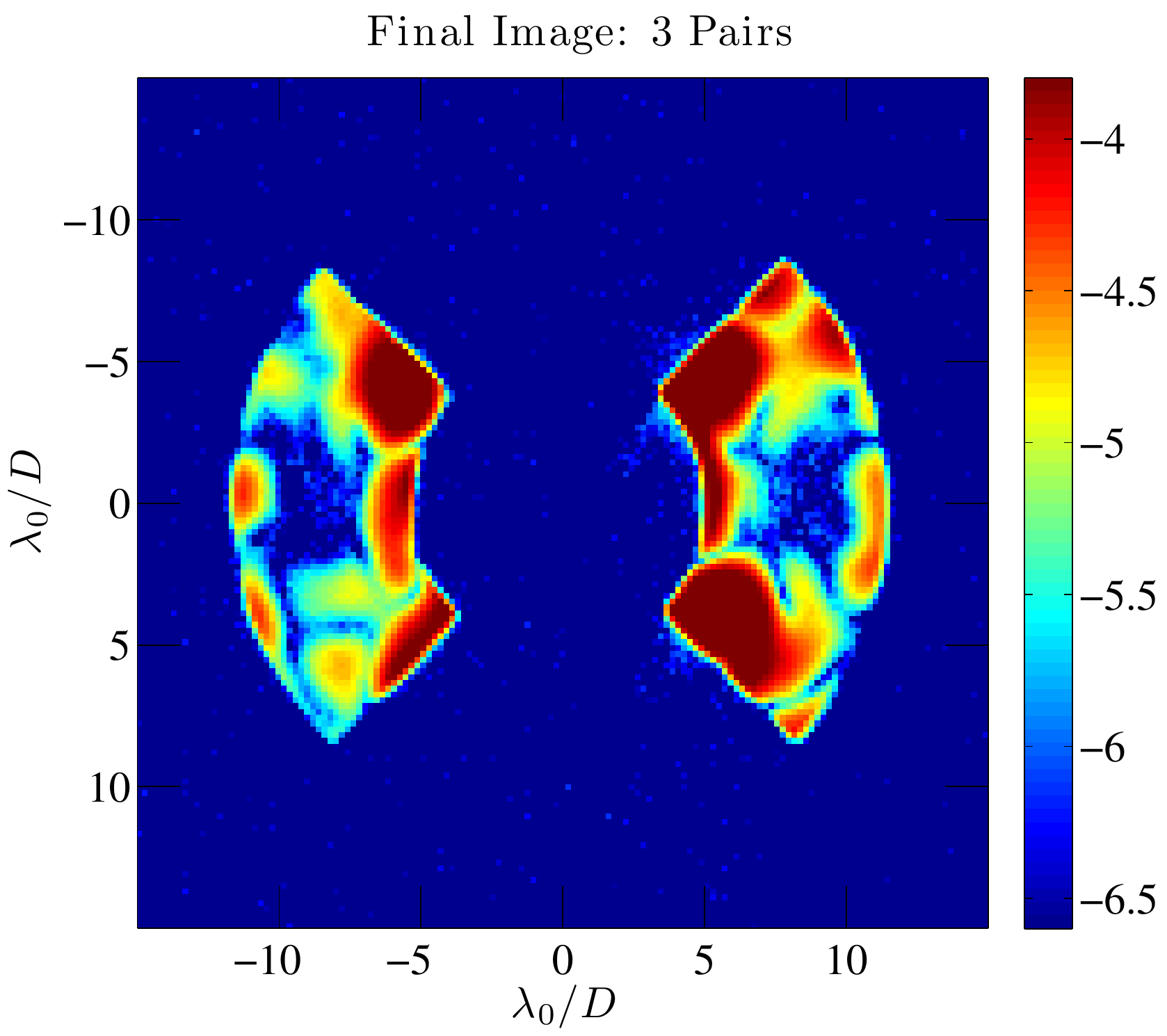}\label{fig:3pairs_final}}
\caption{Experimental results of sequential DM correction using the discrete time extended Kalman filter with 3 image pairs to build the image plane measurement, $z_k$.  The dark hole is a square opening from 7--10 $\times$ -2--2 $\lambda/D$ on both sides of the image plane.  (a) The aberrated image.  (b)  Contrast plot. (c) The corrected image. Image units are log(contrast). }\label{fig:3pairs}
\end{figure}

\begin{figure}[h!]
\centering
\subfigure[]{\includegraphics[width = 0.32\textwidth]{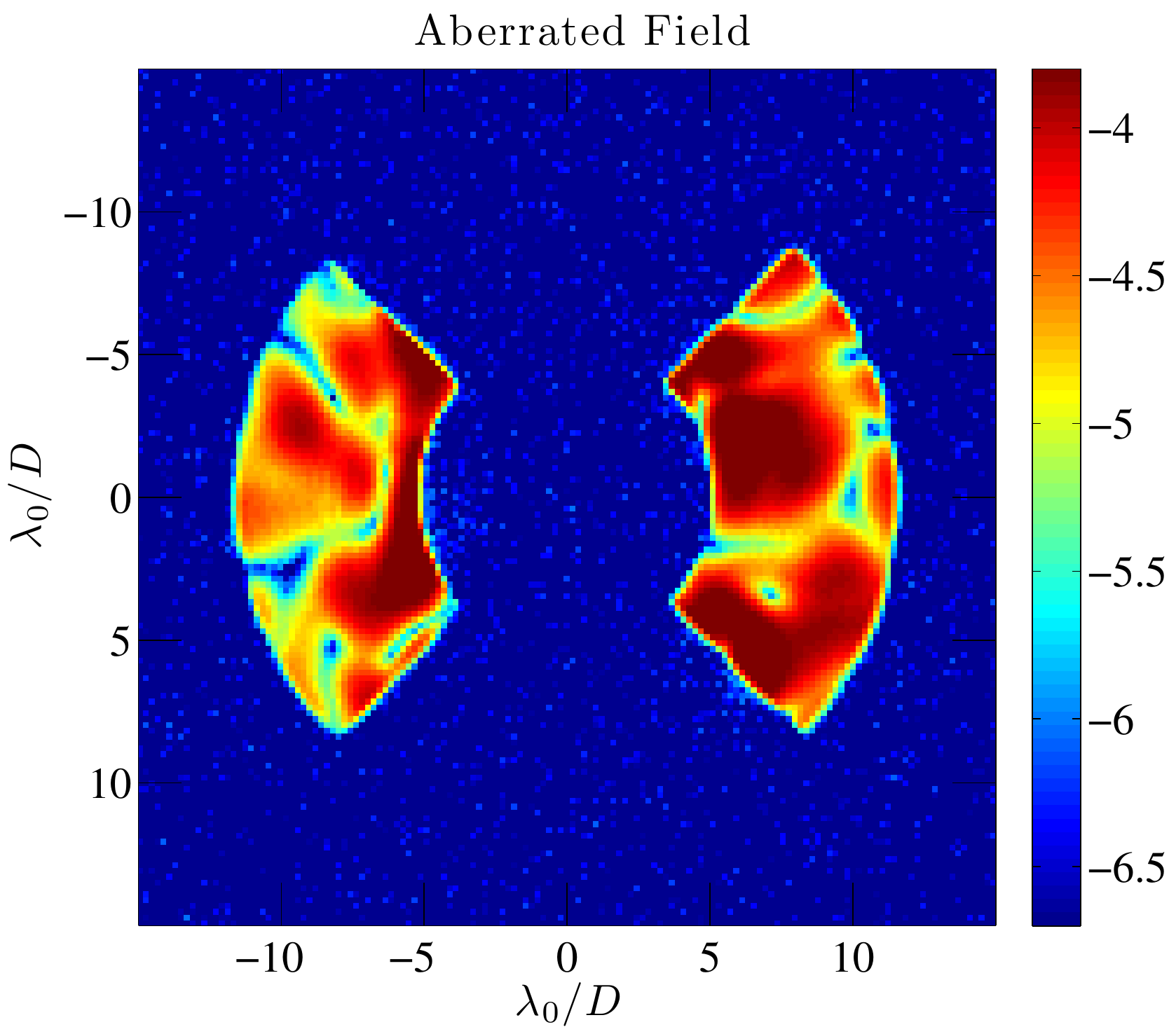}\label{fig:2pairs_initial}}
\subfigure[]{\includegraphics[width = 0.335\textwidth]{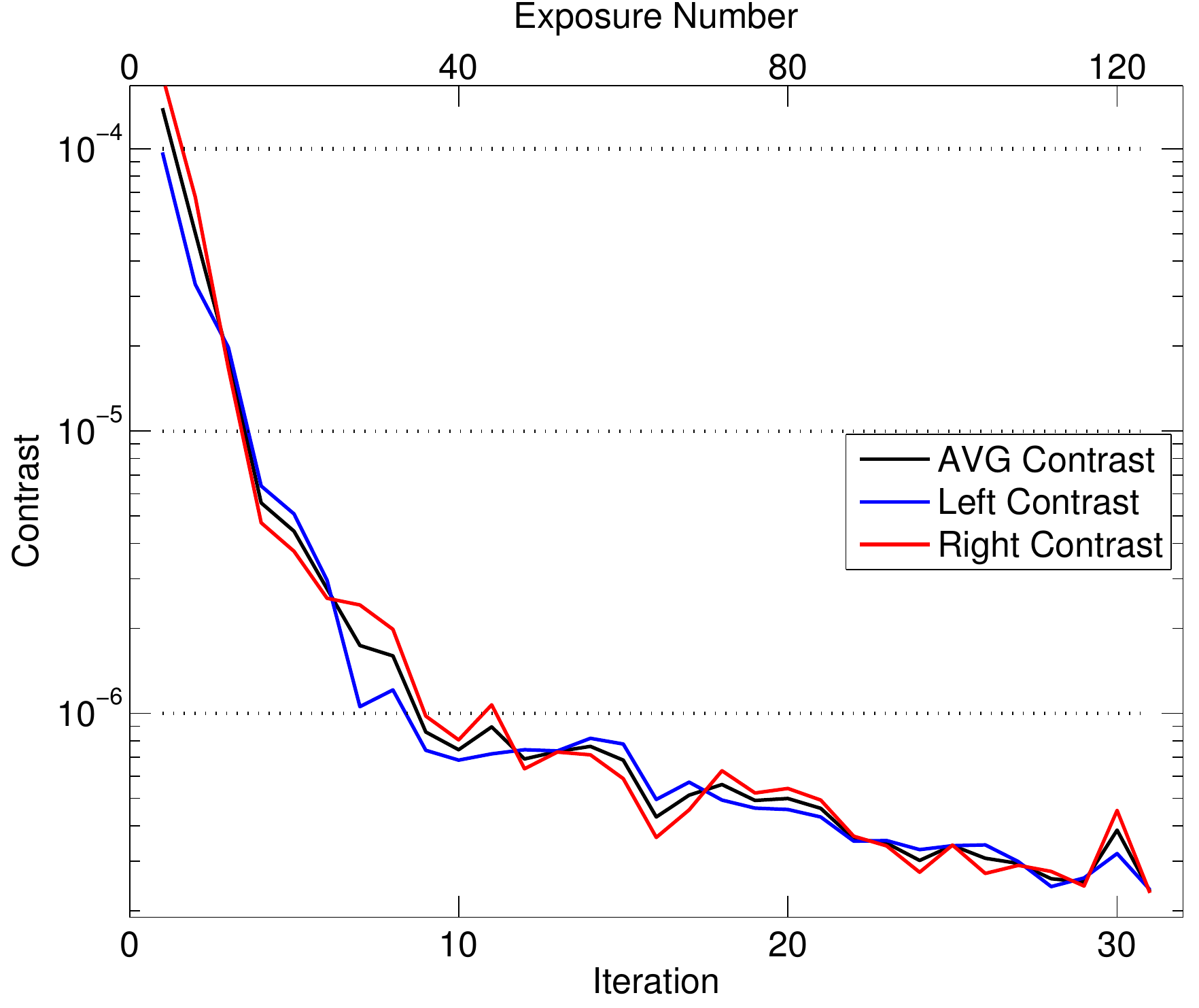}\label{fig:2pairs_contrast}}
\subfigure[]{\includegraphics[width = 0.32\textwidth]{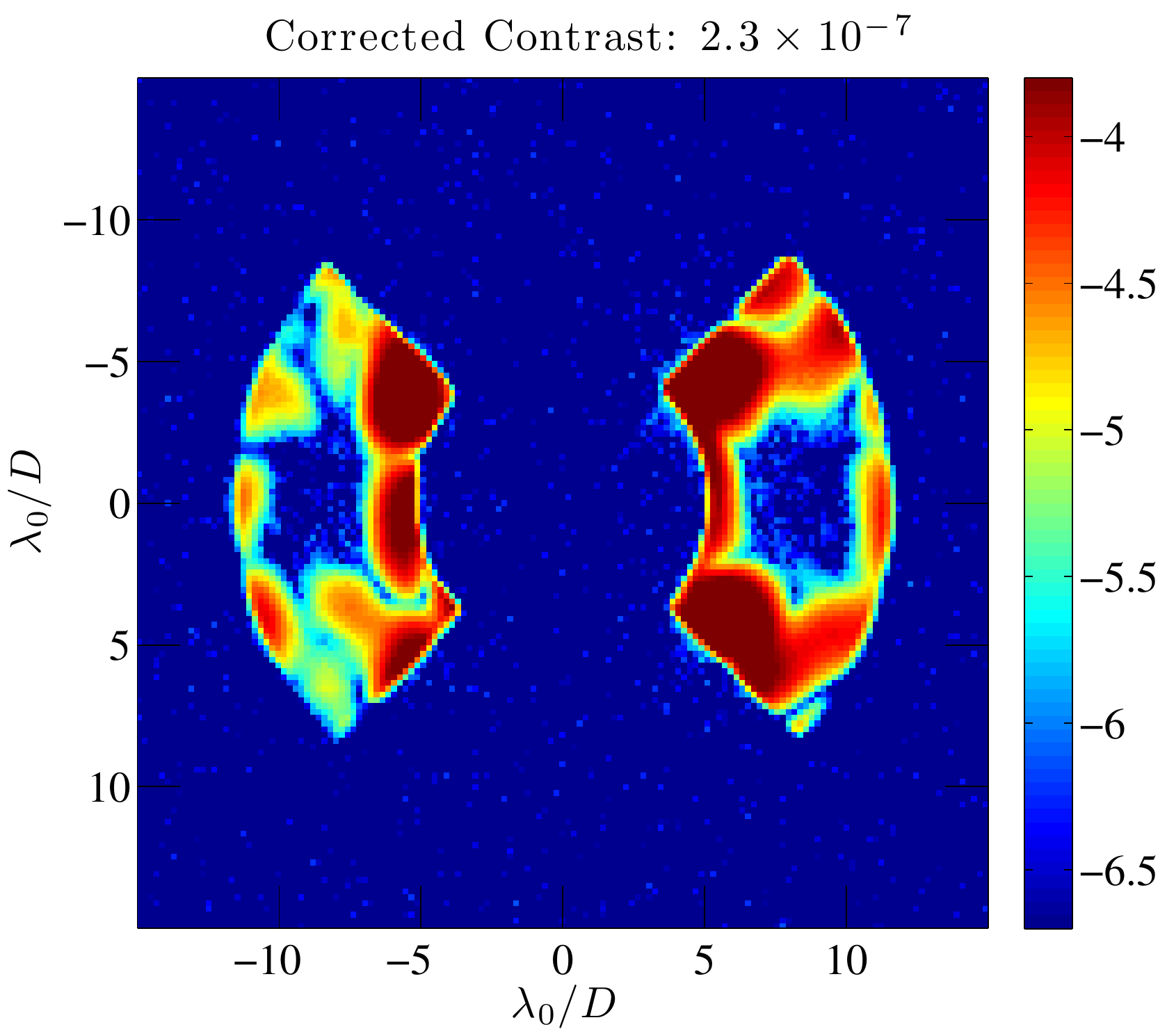}\label{fig:2pairs_final}}
\caption{Experimental results of sequential DM correction using the discrete time extended Kalman filter with 2 image pairs to build the image plane measurement, $z_k$.  The dark hole is a square opening from 7--10 $\times$ -2--2 $\lambda/D$ on both sides of the image plane.  (a) The aberrated image.  (b)  Contrast plot. (c) The corrected image. Image units are log(contrast). }\label{fig:2pairs}
\end{figure}

Reducing the number of image pairs to three, the correction algorithm reached a contrast level of $5.0 \times 10^{-7}$ using only $120$ estimation images, as shown in Fig.\ref{fig:3pairs}.
After fine tuning the covariance and noise matrices, the correction algorithm achieves a contrast of $2.3\times10^{-7}$ in $30$ iterations using only two image pairs per iteration, shown in \fig{fig:2pairs}. The results are better than the three and four pair cases because we improved the covariance initialization and increased the number of numerical iterations ($p$ from \S\ref{sec:iterative}) of the Kalman filter for a single control step. As discussed in \S\ref{sec:iterative},  numerically iterating the filter is critical to its performance, despite the fact that no additional measurements are taken, because it accounts for nonlinearity not captured when numerically propagating the control signal via $\Gamma_{k-1}$.

\begin{figure}[h!]
\centering
\subfigure[]{\includegraphics[width = 0.32\textwidth]{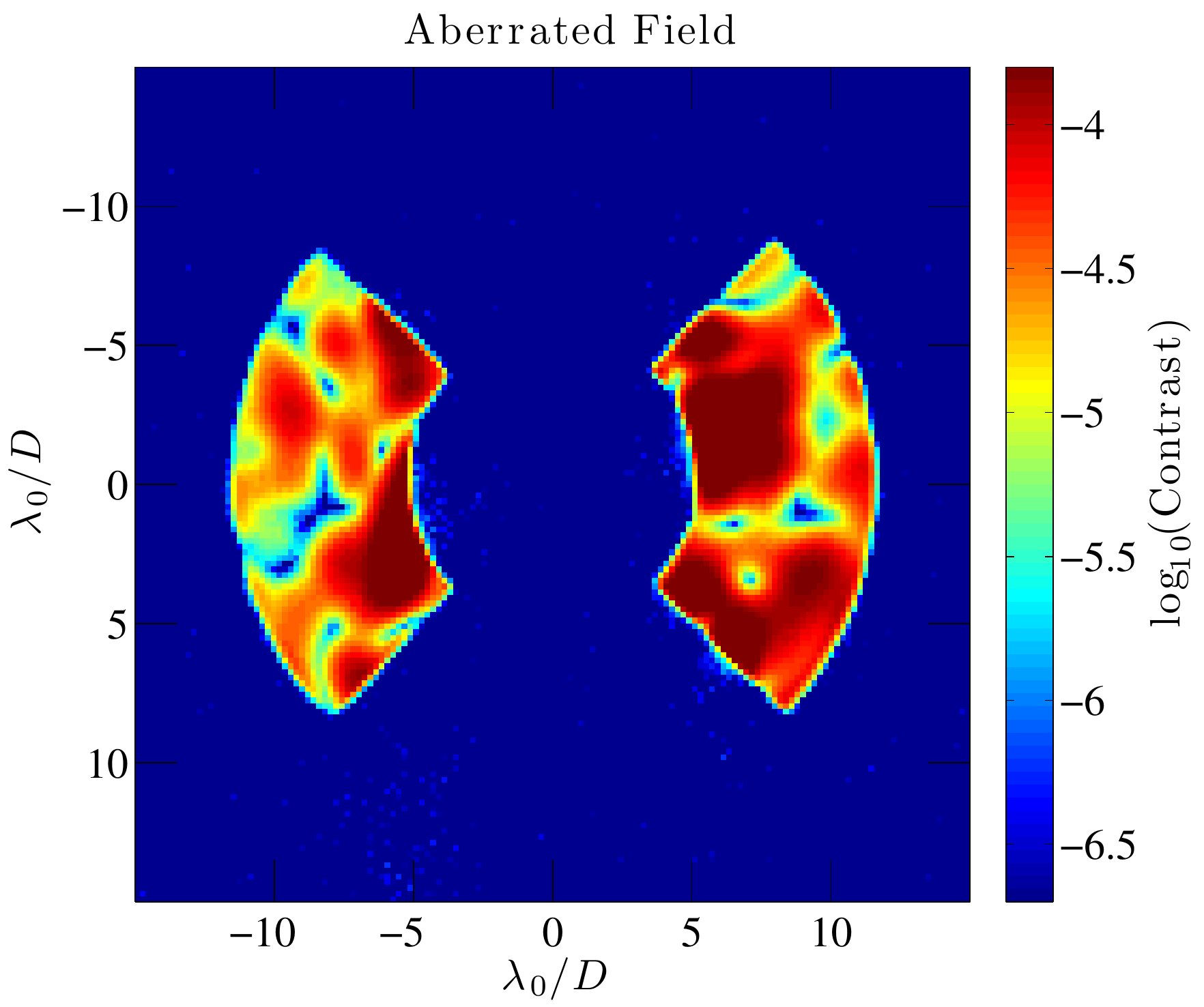}\label{fig:onepair_initial}}
\subfigure[]{\includegraphics[width = 0.335\textwidth]{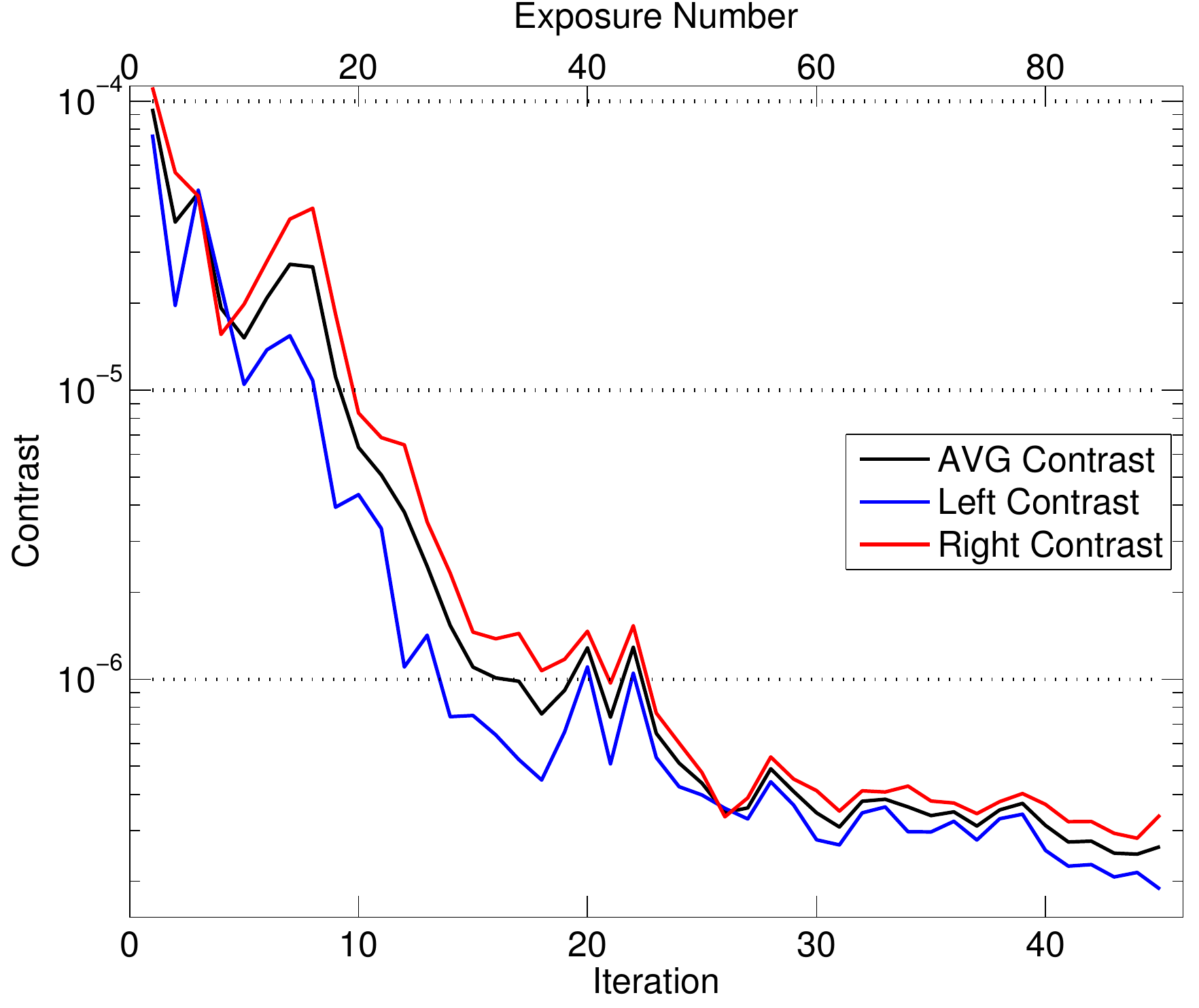}\label{fig:onepair_contrast}}
\subfigure[]{\includegraphics[width = 0.32\textwidth]{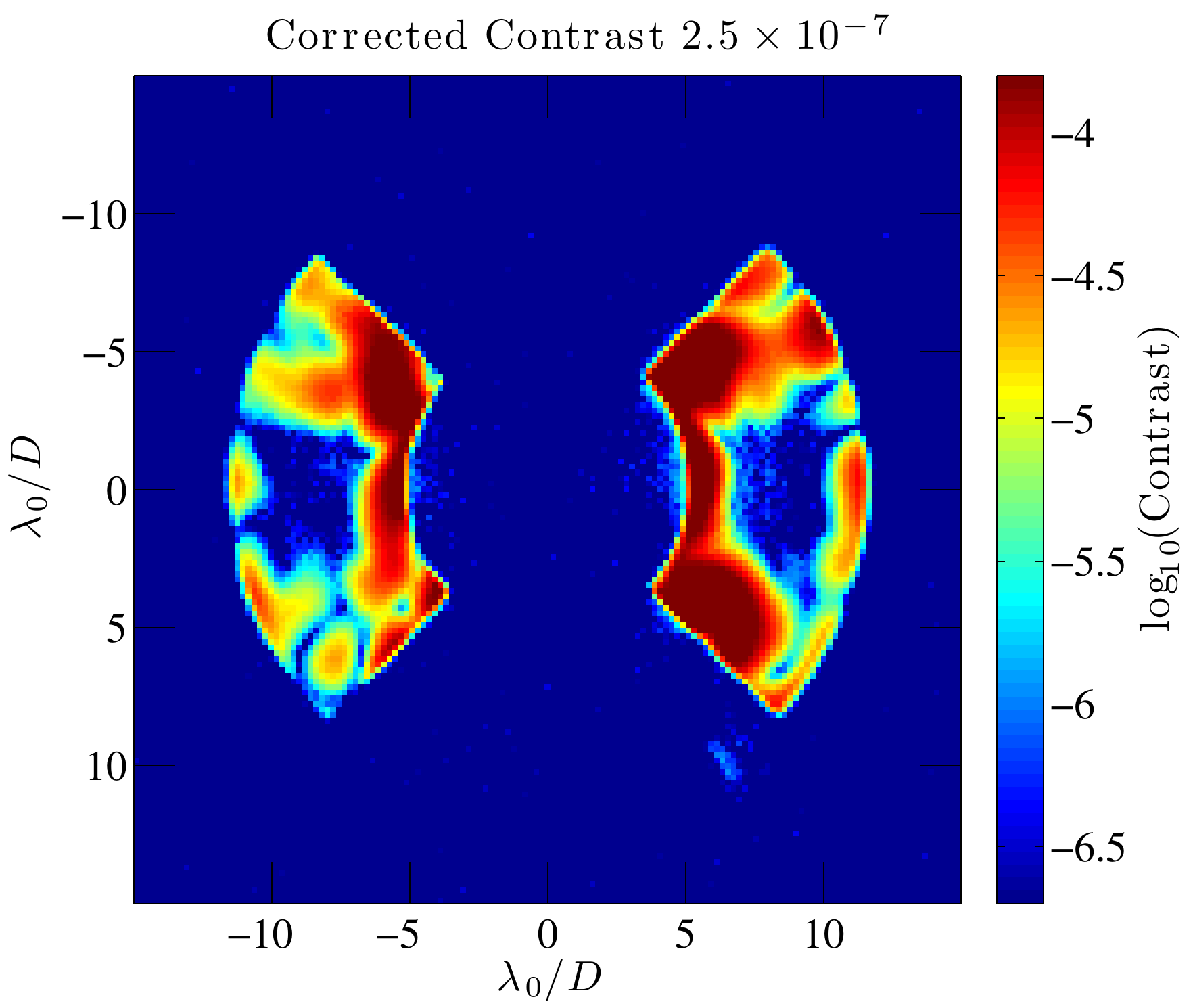}\label{fig:onepair_final}}
\caption{Experimental results of sequential DM correction using the discrete time extended Kalman filter with one image pair to build the image plane measurement, $z_k$.  The dark hole is a square opening from 7--10 $\times$ -2--2 $\lambda/D$ on both sides of the image plane.  (a) The aberrated image.  (b)  Contrast plot. (c) The corrected image. Image units are log(contrast). }\label{fig:1pair}
\end{figure}

Reducing the number of measurements to a single pair we find a very interesting result. The quality of the measurement at any particular time step of the algorithm is now dependent on the quality of that particular probe shape. If a probe does not happen to modulate the field well $H_k$ drops rank, adding noise to the gain computation and estimate update of \eq{eq:gain} and \eq{eq:estupdate}. As a result, the controller can actually degrade the contrast in the dark hole at that time step. Additionally, a single probe may not modulate a specific location of the field well, so we must choose a different probe shape at each iteration to guarantee good coverage of the entire dark hole in closed loop. Starting from an aberrated field with an average contrast of $9.418\e{-5}$, \fig{fig:onepair_initial}, we achieved a contrast of $3.1\e{-7}$ in $30$ iterations and $2.5\e{-7}$ in $43$ iterations of control, \fig{fig:onepair_final}. Looking at the contrast plot in \fig{fig:onepair_contrast}, the sensitivity of a single measurement update to the quality of the probe is very clear. What is interesting, however, is that the modulation in contrast damps out over the control history. While we do not suppress as quickly with regard to iteration we achieve our ultimate contrast levels with fewer total measurements, which is the ultimate performance metric for efficiency of the estimation and control algorithm. Thus, even with one measurement update per iteration the prior history stabilizes the estimate in the presence of the measurement update's poor signal-to-noise at high contrast levels. More encouraging is that these results use the analytical probe shapes described by \eq{eq:probe}. The choice in phase shifts, $\theta_u$ and $\theta_v$, is somewhat arbitrary, with no criteria to evaluate how effectively each modulates the field. In fact, we had to carefully choose our phase shifts to improve the convergence rate of the single probe pair results. If we could choose probe shapes that are guaranteed to modulate the dark hole well, we would see a dramatic improvement in the rate of convergence for a single measurement update.

A very promising aspect of this estimation scheme is that its performance did not degrade as the amount of measurement data was reduced. With only 86 estimation images it was capable of reaching the same final contrast achieved by the DM diversity algorithm in \S\ref{diversity}, 
which required $240$ images to maintain an estimate of the entire control history. 
Thus by making the estimation method more dependent on a model we were able to reduce our need to measure deterministic perturbations in the image plane electric field.

\section{Optimal Probes: Using the Control Signal}\label{sec:optprobes}
The results of \S\ref{sec:results} use the analytical probe shapes described by \eq{eq:probe}. As discussed in \S\ref{sec:results}, the choice of the probe shape used for estimation is critical to create a well-posed problem, and has been somewhat of an art in the high contrast imaging community. These shapes are intended to modulate the field as uniformly as possible so that $H_k$ is full rank. However, no formalism has been proposed to determine the ``best" shapes to probe the dark hole. In any dark hole there are discrete aberrations that are much brighter than others, requiring that we apply more amplitude to those spatial frequencies. Conversely the bright speckles raise the amplitude of the probe shape, making it too bright to measure the dim speckles with an adequate SNR. Excluding this issue, we also cannot truly generate the analytical functions, as the actual surface is a sum of individual influence functions. Even the DM with the highest actuator density available, the Boston Micromachines 4K-DM\copyright, can only approximate each function with 64 actuators. We account for the true shape in our model but this shape does not truly probe each pixel in the dark hole with equal weight, which was the primary advantage of using the analytical function for a probe shape in the first place. Fortunately, we can once again appeal to our mathematical model for estimation and control to help determine an adequate probe shape. In closed loop, the control law incorporates the influence function model of the surface to determine a shape that necessarily modulates the aberrated field in the dark hole. In principle, if we apply the conjugate of the control shape we will increase the energy of the aberrated field. Instead of applying probes in addition to the control shape, we use the control shape itself (and its negative) to probe the elecric field in the dark hole. Thus, we can rely on the controller to compute optimal probe shapes that will inherently modulate brighter speckles more strongly than dimmer speckles and we have confidence that the shape will adequately perturb the dark hole field.
\begin{figure}[h!]
\centering
\subfigure[]{\includegraphics[width = 0.32\textwidth]{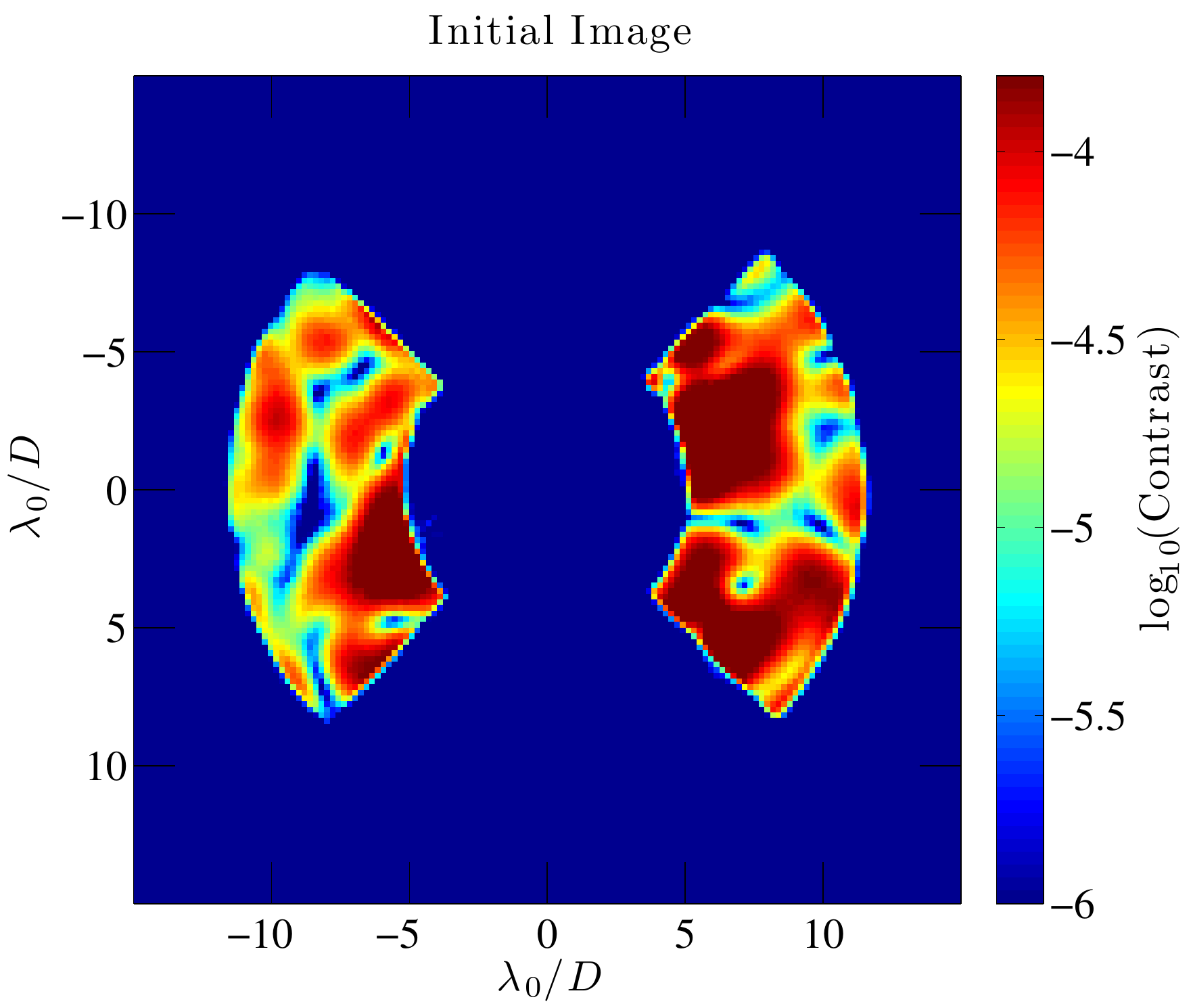}\label{fig:dmprobe_initial}}
\subfigure[]{\includegraphics[width = 0.335\textwidth]{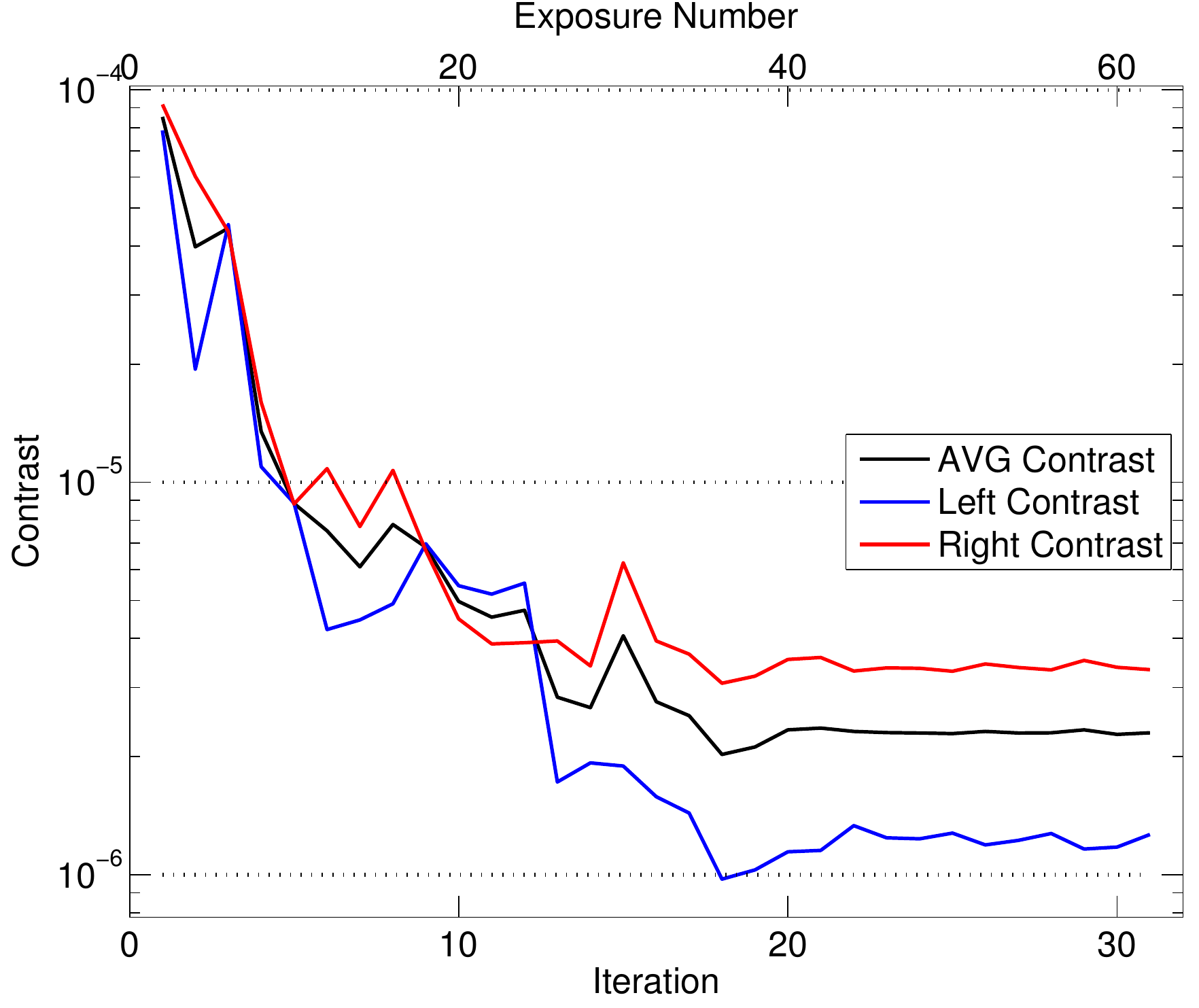}\label{fig:dmprobe_contrast}}
\subfigure[]{\includegraphics[width = 0.32\textwidth]{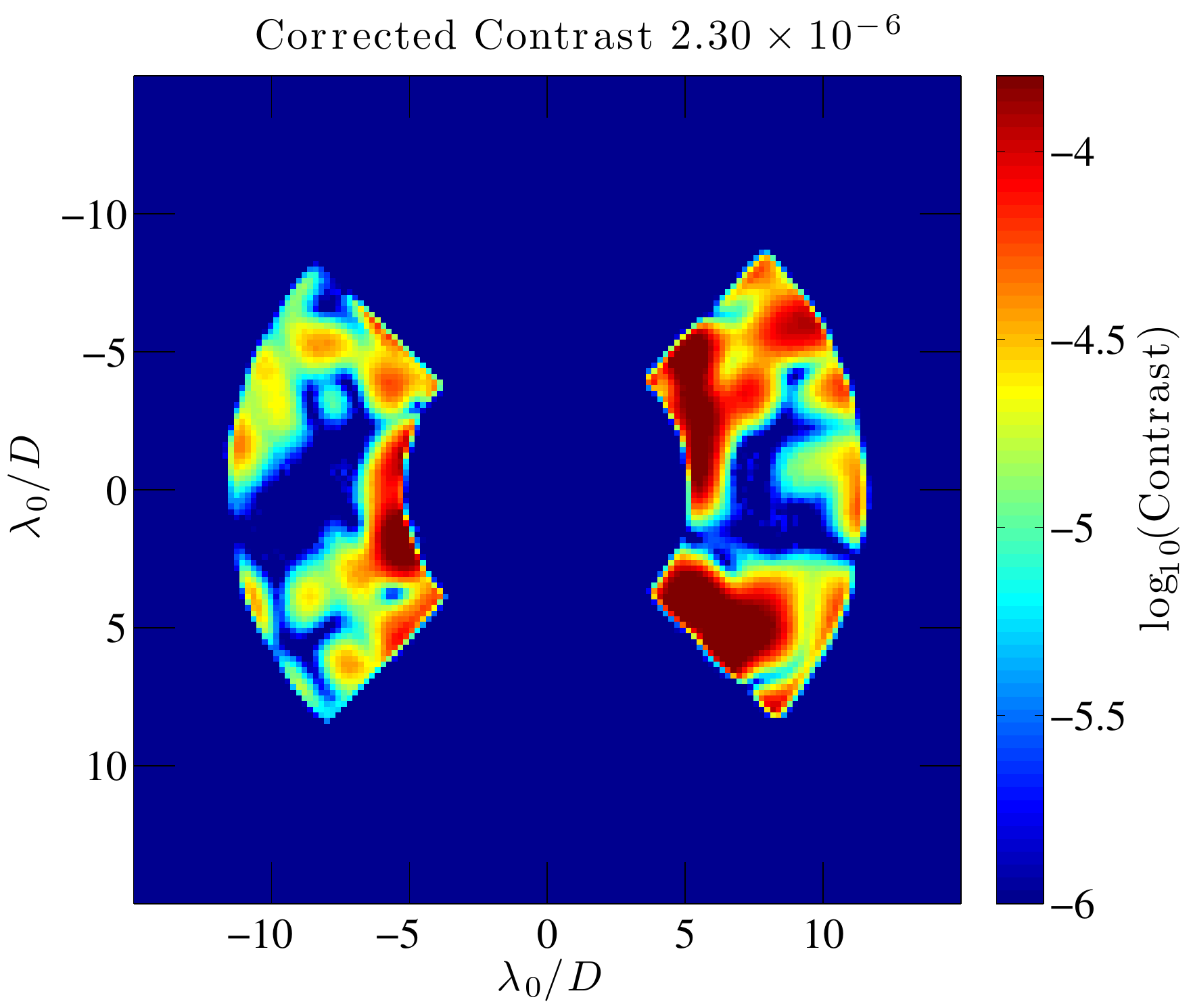}\label{fig:dmprobe_final}}
\caption{Experimental results of sequential DM correction using the discrete time extended Kalman filter with the control effect and it's conjugate shape used as the only probe pair for the measurement update, $z_k$.  The dark hole is a square opening from 7--10 $\times$ -2--2 $\lambda/D$ on both sides of the image plane.  (a) The aberrated image.  (b)  Contrast plot. (c) The corrected image. Image units are log(contrast). }\label{fig:probepair}
\end{figure}

\fig{fig:probepair} shows preliminary results using the control and its conjugate shape as the probe pair for estimating the field with the Kalman filter. The current contrast level is limited to $2.30\e{-6}$ on both sides of the image plane. The asymmetry of the dark holes actually hints at the reason for not achieving a higher contrast level. In a 2-DM system, the idea that the optimal control signal can be used as the probe shape requires that we use both mirrors to probe the field. Since the estimator is currently written assuming one mirror is probing the field, we were forced to collapse both control shapes onto the same DM. This means that we are not fully perturbing the field as we would have expected, leaving a particular set of aberrations unprobed. This is not a problem for a single sided dark hole using one DM, where the control and probe surfaces are one and the same. For a two DM system, we simply need to reformulate the estimator as a function of both DMs for this concept to be effective.
\section{Conclusions}\label{conclusions}
We have demonstrated a discrete time iterative Kalman filter to estimate the image plane electric field in closed loop. The experimental results only required a single linearization for the estimator, and we have mathematically demonstrated a suitable extension to an IEKF in the event that we must re-linearize about a new DM shape. This type of progress is critical for improving the efficiency and robustness of future coronagraphic missions, thereby maximizing the likelihood of planetary detections. We demonstrate the fastest suppression to date in the Princeton HCIL, only requiring a single measurement at each iteration. This represents a $\approx$70\% reduction in the number of images required to estimate the image plane electric field. Having increased the speed of the estimator, this makes it a feasible focal plane estimation technique for ground-based coronagraphic instruments. The closed loop nature of the estimator is attractive because it also stabilizes the estimate to measurement noise. So long as the probe adequately modulates the intended dark hole area, a measurement update with poor signal-to-noise does not adversely affect the covariance of the state estimate. Having controlled the field with only a single measurement update, we have demonstrated that not all probe shapes are best for estimation, motivating us to attempt using the control shape as the probing function. Preliminary results for using the control shape as a probe signal are promising, and may further reduce the number of required exposures to one image per iteration. The Kalman filter also opens up the possibility of adaptive control techniques to learn laboratory physical parameters, sensor fusion concepts for integration with extreme adaptive optics systems, and bias estimation to gain certainty in planetary detection using only the control history.

\section*{Acknowledgements}
This work was funded by NASA Grant \# NNX09AB96G and the NASA Earth and Space Science Fellowship

\bibliographystyle{osajnl}

\bibliography{bibLibrary}

\end{document}